\documentclass[galaxies,review,accept,moreauthors,pdftex]{Definitions/mdpi}

\firstpage{1} 
\pubvolume{0}
\issuenum{0}
\articlenumber{0}
\pubyear{2022}
\copyrightyear{2020}
\datereceived{2022/06/21} 
\dateaccepted{Under consideration} 
\datepublished{Under consideration} 
\hreflink{https://doi.org/} 

\newcommand{\tview}{$\theta_{\rm v}$}
\newcommand{\respone}[1]{#1}

\Title{The Structure of Gamma Ray Burst Jets}

\TitleCitation{Title}

\Author{Om~Sharan Salafia$^{1,2,3,\dagger,*}$\orcidA{} and Giancarlo Ghirlanda$^{3,2,\dagger}$\orcidB{}}

\AuthorNames{Giancarlo Ghirlanda and Om~Sharan Salafia}

\AuthorCitation{Salafia, O.; Ghirlanda, G.}

\address{%
$^{1}$ \quad Università degli Studi di Milano-Bicocca, Dipartimento di Fisica ``G.\ Occhialini'', piazza della Scienza 3, I-20126 Milano (MI),  Italy\\
$^{2}$ \quad INFN -- Sezione di Milano-Bicocca, piazza della Scienza 3, I-20126 Milano (MI),  Italy\\
$^{3}$ \quad INAF -- Osservatorio Astronomico di Brera, via E.~Bianchi 46, I-23807 Merate (LC), Italy}

\corres{Correspondence: om.salafia@inaf.it}

\firstnote{These authors contributed equally to this work.}

\abstract{Due to relativistic bulk motion, the structure and orientation of gamma-ray burst (GRB) jets have a fundamental role in determining how they appear. The recent discovery of the GW170817 binary neutron star merger and the associated GRB boosted the interest in the modelling and search of signatures of the presence of a (possibly quasi-universal) jet structure in long and short GRBs. In this review, following a pedagogical approach, we summarize the history of GRB jet structure research over the last two decades, from the inception of the idea of a universal jet structure to the current understanding of the complex processes that shape the structure, that involve the central engine that powers the jet and the interaction of the latter with the progenitor vestige. We put some emphasis on the observable imprints of jet structure on prompt and afterglow emission and on the luminosity function, favoring intuitive reasoning over technical explanations. 
}


\begin{document}

\section{Introduction}

Jets, in the form of collimated outflows of plasma possibly endowed with magnetic fields, are ubiquitous in astrophysics. They typically extend over orders of magnitude in distance from their birthplace (from parsec scales in protostars to $>$kpc scales in galaxies hosting supermassive black holes), in redshift (with jet signatures being detected in association to the most distant galaxies known so far up to $z\sim9$) and in luminosity, reaching the largest values in gamma-ray bursts (GRBs). 

GRBs are luminous, extra-galactic transients powered by compact objects (black holes -- BH, neutron stars -- NS) produced by the core-collapse of a massive stars or by the merger of a compact object binary (most likely NS-NS or NS-BH). In the most widely accepted scenario, the `central engine' (that is, the system consisting of the compact object and possibly a surrounding accretion disk) launches a bipolar relativistic collimated outflow. Bulk energy dissipation in such outflow produces a bright, highly variable, non-thermal `prompt' emission in the X-ray/$\gamma$-ray band. The outflow deceleration by the external circum-burst medium produces the long-lasting multi-wavelength `afterglow' emission extending from the $\gamma$-rays through the optical to the radio band. 

\begin{figure}
    \centering 
    \includegraphics[width=0.7\textwidth]{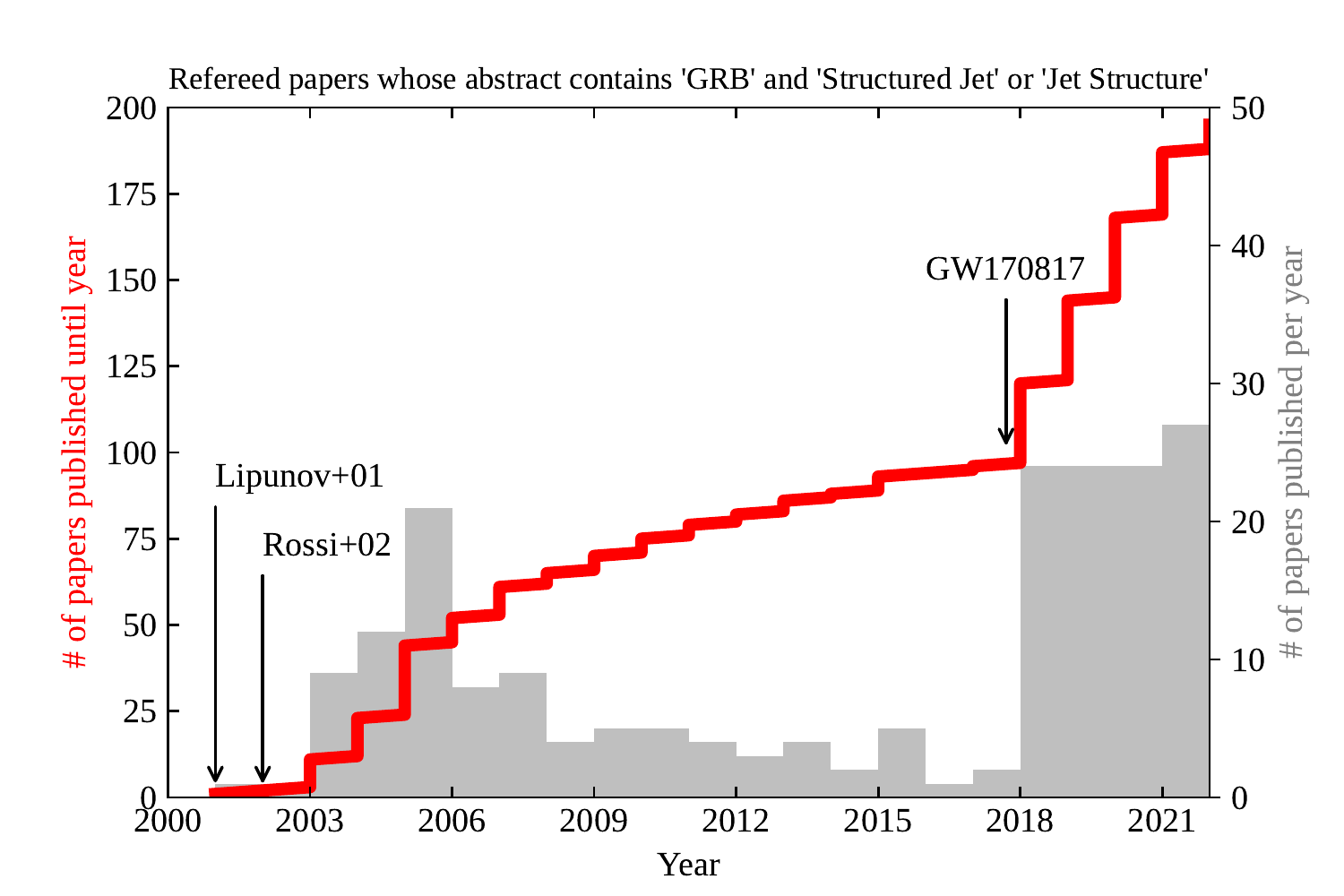}
    \caption{Timeline of scientific papers about GRB structured jets. The red solid line shows the cumulative number (shown on the left-hand axis) of refereed papers that contain ``gamma-ray burst'' and ``structured jet'' or ``jet structure'' in their abstract, according to the NASA ADS \cite{ADSNASA}. The grey histogram (number shown on the right-hand axis) shows the corresponding number of papers published per year. The dates of the two seminal papers \cite{Lipunov2001,Rossi2002} and of the GW170817 discovery \cite{Abbott2017PRL,Abbott2017ApJ,Abbott2017-mm} are annotated. \respone{We note that alternative nomenclatures with respect to the `structured jet'/`jet structure' used here exist, hence the actual number of papers on the subject could be higher.}}
    \label{fig:paper_history}
\end{figure}

The presence of {\it relativisic} outflows in GRBs is supported by some theoretical arguments and a few compelling observational evidences. The very fast prompt emission light curve variability requires the source to be very compact, but the observation of non-thermal prompt emission spectra extending above MeV photon energies indicates that the source is optically thin to pair production by photon-photon annihilation. This apparent contradiction can hardly be reconciled without invoking highly relativistic expansion, which eases the constraints by both decreasing the comoving photon energy by a $\Gamma$ factor (the bulk expansion Lorentz factor) and increasing the source size limit imposed by variability by a $\Gamma^2$ factor (e.g.\ \cite{Cavallo1978,Paczynski1986-vi,Goodman1986,Krolik1991-la,Lithwick2001-sc}). Even more directly, the apparent size increase of $\sim$0.3 pc in $\sim$50 rest-frame days as measured for the first time in the nearby GRB~030329 \citep{Taylor2004-zo} suggested an apparently superluminal expansion speed, indicating relativistic bulk motion \citep{Rees1966}. {\it Collimation} of GRB outflows is required to reduce the otherwise huge $\gamma$-ray isotropic equivalent\footnote{This nomenclature refers to the energy and/or luminosity of a GRB computed assuming isotropic emission. Because of relativistic `beaming' (i.e.\ aberration) of radiation, the vast majority of the observable photon flux comes from emitting regions moving within a tiny $1/\Gamma$ angle around the line of sight, making an isotropic outflow (and hence isotropic emission) essentially indistinguishable from one that expands radially within a $\theta_\mathrm{j}\gtrsim 1/\Gamma$ collimation angle \cite{Rhoads1997-yh}.} energy reaching $E_{\gamma,\rm iso}\sim 10^{54-55}$ erg (e.g.\ GRB~990123 \cite{Briggs1999-zl} and GRB~130427A \cite{Preece2013-dv}), which would require the mind-boggling conversion of 1--5 $M_{\odot}$ rest mass energy into $\gamma$-rays with 100\% efficiency without invoking collimation. If the outflow is collimated within an angle $\theta_\mathrm{j}$, such energy budget is reduced by a `beaming' factor $f_\mathrm{b}=(1-\cos\theta_{\rm j})\sim \theta_\mathrm{j}^2/2 \approx 0.004 (\theta_\mathrm{j}/5^\circ)^2$. The collimation angle $\theta_\mathrm{j}$ is typically estimated from a steepening of the afterglow light curve around a few days after the initial gamma-ray burst, interpreted as the signature of the presence of a jet \citep[][see also \S\ref{sec:late_afterglow}]{Rhoads1997-yh,Rhoads1999,Fruchter1999-hv,Kulkarni1999-hd,Stanek1999-wo,Harrison1999-ge}. Such a feature, often referred to as a `jet break', arises as the relativistic beaming angle $1/\Gamma$ (which increases during the afterglow phase due to the deceleration of the blastwave, i.e.\ the expanding shock produced as the jet expands within the external medium) becomes comparable to $\theta_{\rm j}$ \cite{Rhoads1997-yh}, allowing the observer to `see' the jet borders. \respone{Typical} collimation angles estimated from the observation of jet breaks range\footnote{\respone{Opening angles as small as $\theta_\mathrm{j}<1^\circ$ have been reported in some studies \citep[e.g.][]{Chandra2012,Lu2012}. We caution that, while opening angles as small as these are not impossible in principle, these estimates are based on assumptions on the interstellar medium density and prompt emission efficiency, and they rely on the interpretation of an observed steepening in the afterglow light curve as a jet break. For the latter intepretation  to hold, the steepening must be achromatic, i.e.\ it must show up independently of the observing band, but it is often impossible to verify it due to the absence of multi-wavelenght observations at the relevant time. For these reasons, such estimates must be taken with a grain of salt.}} from $\theta_\mathrm{j}\sim 4^\circ$ in `long' GRBs \cite{Frail2001-ne} to $\theta_\mathrm{j}\sim 16^\circ$ in `short' GRBs \cite{Fong2015-ya}.  

For simplicity, the jet of GRBs has been long modelled as a conical outflow with a constant energy per unit solid angle $\mathrm{d}E/\mathrm{d}\Omega(\theta)$ and bulk Lorentz factor $\Gamma(\theta)$ within its aperture $\theta\leq \theta_\mathrm{j}$ (here $\theta$ is the angle from the jet symmetry axis). This basic model is typically referred to as the `uniform', `homogeneous' or `top-hat' jet structure model. If the jet is observed within $\theta_\mathrm{j}$, the steepening in the afterglow light curve is used to infer the jet opening angle $\theta_{\rm j}$ from where the true burst energy can be derived $E_{\gamma}\sim E_{\gamma,\rm iso}\theta_{\rm j}^2/2$. It was found by \cite{Frail2001-ne,Berger2003-az,Ghirlanda2004-ny} that $E_{\gamma}$ is narrowly distributed around $10^{51}$ erg, suggesting a standard energy reservoir in GRBs. Within a `top-hat' jet model, this implies that $E_{\gamma,\mathrm{iso}}$ scales as $\theta^{-2}$. On the other hand, some authors \cite{Lipunov2001,Rossi2002,Zhang2002-sz} soon realized that the same observations could be explained assuming GRBs jets to posses a universal {\it structure} $\mathrm{d}E/\mathrm{d}\Omega\propto \theta^{-2}$ (see \S\ref{sec:definition_jet_structure} for a precise definition).

The evolving interest in the structure of GRB jets can be seen in Fig.~\ref{fig:paper_history}, where we have collected from the NASA ADS all the papers mentioning ``gamma-ray burst'' and ``structured jet'' or ``jet structure'' in their abstracts. The red line is the cumulative distribution of the grey histogram and shows two clear ``steps'': an initial growing interest in structured jets corresponding to the 2000-2006 period and a recent ``explosion'' of interest prompted by the discovery and interpretation of the gamma-ray burst GRB~170817A associated to the first binary neutron merger gravitational wave event \cite{Abbott2017PRL,Abbott2017ApJ,Abbott2017-mm}.

The initial interest in structured jets in the early 2000s was in part driven by attempts at explaining the diversity of GRB energetics within a unifying scenario where all jets share a universal structure.
Two analytical functions were explored initially to describe the jet structure: a power law jet with $\mathrm{d}E/\mathrm{d}\Omega\propto \theta^{-2}$, as suggested by the $E_\gamma$ clustering described above and supported by early analytical studies \cite{Lazzati2005-ar} and numerical simulations \cite{Morsony2007} of the jet emerging from its progenitor star envelope (see \S\ref{sec:stages}); 
a Gaussian jet with $\mathrm{d}E/\mathrm{d}\Omega\propto \exp(-(\theta/\theta_{\rm c})^2/2)$ \cite{Rossi2002,Zhang2002-sz,Granot2003,Kumar2003}, where most of the jet energy is contained within two times the `core' opening angle $\theta_{\rm c}$, which is a more realistic representation of a nearly sharp-edged jet. Less continuous structures, such as one composed by two nested uniform jets (a narrow, fast and energetic jet surrounded by a wider, slower and weaker layer\footnote{Notably, a two-component jet structure has also been proposed to interpret jets in radio galaxies \cite{Ghisellini2005-hk}.}) were considered, motivated by the possibility to explain the optical afterglow bumps observed in a few GRBs \cite{Berger2003-az,Racusin2008-kq}. 

In the same period many attempts were made at identifying, in the observational data then available, distinctive features of a structured jet. Modelling of the afterglow light curve of GRB030329 \cite{Van_der_Horst2005-oz} suggested a structured jet as a viable interpretation of the low frequency data, although alternative interpretations were not excluded.  The sharpness of the light curve change across the jet break time, which in the structured jet scenario provides a measure of the viewing angle \tview\ \cite{Rossi2002}, depends on the jet structure and on the viewing angle, with sharper breaks corresponding to larger \tview\ \cite{Granot2005-kv}.  However, also the jet lateral expansion affects the shape of the light curve around the jet break time \cite{Salmonson2003-nf}. Attempts at testing the universal jet structure model \cite{Perna2003-tj,Dai2005-xm,Lloyd-Ronning2004-dm} were mainly limited by the few events with measured redshifts and jet breaks \cite{Nakar2004-sf}. Linear polarization measurements of the afterglow emission were also considered as diagnostics for the jet structure \cite{Ghisellini1999,Rossi2004}, despite the polarization depends also on the configuration of the magnetic field in the emission region \cite{Lazzati2004-xt, Granot2005-kv,Granot2021Galax...9...82G}. Considerably different rates of GRB afterglows without a corresponding prompt emission detection (so called orphan afterglows) are predicted in the case of a structured jet with respect to the conical uniform scenario \cite{Totani2002,Granot2002_jetted_afterglow,Nakar2002,Rau2006,Perna1998,Rossi2008-mh,Ghirlanda2014,Ghirlanda2015}. 

Owing to the difficulties in identifying distinctive signatures in the available data of the structured jet scenario (see \cite{Granot2007}), the community started to loose interest in it during the 2006-2017 period. 
The discovery of GRB~170817A \citep{Abbott2017ApJ,Goldstein2017,Savchenko2017} associated with the GW170817 gravitational wave source \citep{Abbott2017PRL} suddenly changed everything (see also \S\ref{sec:GW170817}): after more than six months of monitoring of the puzzling non-thermal afterglow of GRB~170817A, a structured jet appeared as the only scenario able to provide a self-consistent interpretation of the shallow evolution of the afterglow light curves \cite{Lamb2017-ih,DAvanzo2018-ex,Alexander2018,Margutti2018-qf,Troja2018-ua,Lazzati2018-vl,Resmi2018-kj,Wu2018,Ryan2020} and of the proper motion \cite{Mooley2018-xr} and small size \cite{Ghirlanda2019} of the very long baseline interferometry (VLBI) images of the source \respone{(see \cite{Margutti2021} for a review of the multi-messenger aspects of GW170817, and \cite{Nakar2020} for a more general review of electromagnetic counteparts of compact binary mergers)}.

\begin{figure}
    \centering
    \includegraphics[width=0.7\textwidth]{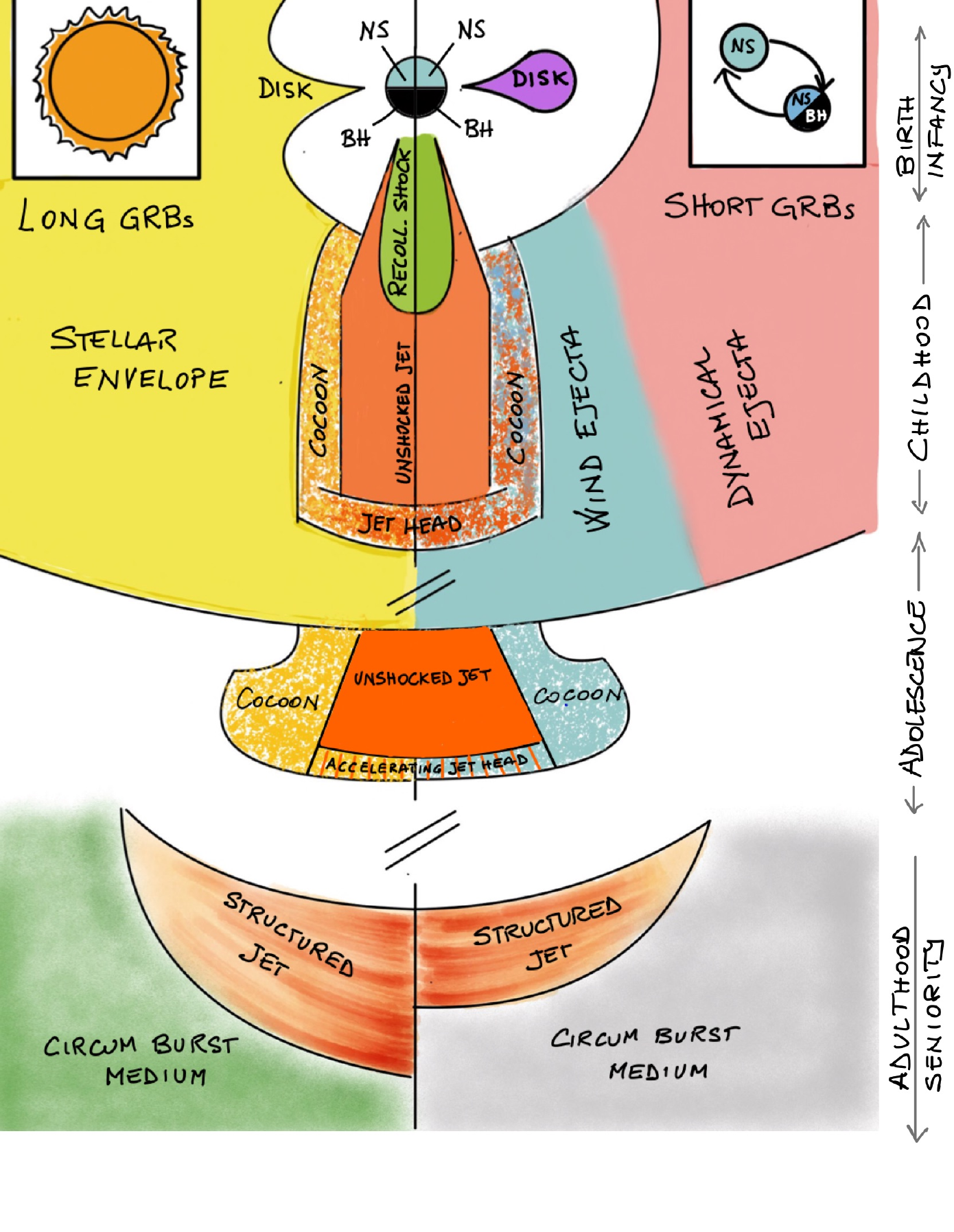}
    \caption{Artist's impression of the different phases in the early evolution of a GRB jet (long and short GRBs schematically represented on the left-hand and right-hand side respectively). The putative progenitors are depicted in the stamps on the top corners. The formation of a compact central object (BH or NS) accompanied by an accretion disk powers a relativistic jet, determining the properties at its base (``birth''). The jet expands within the progenitor vestige (stellar envelope -- left -- or merger ejecta and disk winds -- right) and starts interacting with it (``infancy''). The cocoon formed by ambient and jet shocked material produces an inward pressure that collimates the jet as it propagates (''childhood''). As the jet breaks out (``adolescence'') its head accelerates and the cocoon blows out. The subsequent phases (``adulthood and seniority'', described in the text) are responsible for the prompt and afterglow emissions. See \S2.2 for the description of the phases thorugh which a jet evolves.}
    \label{fig:disegno} 
\end{figure}

Why is the structure important? The structure determines the properties of the emission for different observers, therefore determining in part the distribution of observable properties and the detectability of these sources. The jet structure carries information about the processes that shape it (the jet-launching mechanism and the interaction between the jet and the ambient medium surrounding the central engine) and is therefore an indirect probe of otherwise unobservable phenomena. Several works developing the concept of the jet structure, its origin, and how it determines the observed properties of GRBs appeared in the literature in the last five years. The presence of a jet with some structure appears unavoidable, considering the phases following the formation of the central engine and, therefore, a growing part of the community is starting to systematically consider GRB observations under this more realistic assumption when interpreting both their prompt and afterglow emission components. However, often the available observations are insufficient to allow for distinguishing between a structured jet from a less realistic assumption of a uniform jet. Most likely, the combination of several observables and the further development of numerical simulations will lead to constrain the structure of GRB jets in the near future. 

The scope of this review is that of introducing the general definition of jet structure (\S2.1) and present a very intuitive description of how the jet acquires its angular structure (\S2.2). A very simplified overview of the mechanisms responsible for the jet launch (\S2.3) and for its propagation up to where it can freely expand (\S2.4) is provided. The possible signatures of the presence of a structured jet on the observed properties of the prompt GRB emission are presented  in \S3. The afterglow emission from a structured jet considering different possible structures and its dependence on the key jet structure parameters is summarized in \S4. \respone{Finally, in \S\ref{sec:GW170817} we briefly review the observations of the non-thermal electromagnetic counterparts of GW170817 and their interpretation in the structured jet scenario.}

\section{Origin of the jet structure}

The mechanisms that shape the structure of a gamma-ray burst relativistic jet are complex and not entirely understood. In this section, we will summarize some general ideas about these processes in a pedagogical manner.

\subsection{General definition of jet structure}\label{sec:definition_jet_structure}

At a fixed time $t$, assuming axisymmetry, radial expansion and a relativistic equation of state $p=e_\mathrm{int}/3$ (where $p$ is the pressure and $e_\mathrm{int}$ is the internal energy density, both measured in the comoving frame of the fluid) for simplicity, the jet structure can be described by four functions of the radial coordinate $r$ and polar angle $\theta$ (measured from the jet axis), namely the modulus of the four-velocity $\Gamma\beta = u(r,\theta,t)$ (where $\Gamma$ is the bulk Lorentz factor and $\beta=(1-\Gamma^{-2})^{1/2}$), the comoving rest-mass density $\rho'(r,\theta,t)$, the dimensionless enthalpy $h(r,\theta,t) = 1 + 4e_\mathrm{int}/3\rho'c^2$ and the magnetization $B^2/4\pi \rho'c^2=\sigma(r,\theta,t)$ (where $B$ is the comoving magnetic field strength, assumed transverse with respect to the expansion). Often, it is possible to limit the discussion to cold ($h\sim 1$) and highly relativistic ($u\sim \Gamma$) parts of the fluid, in which case the rest-mass density, magnetization and Lorentz factor $\Gamma(r,\theta,t)$ are sufficient. If the radial structure is unimportant and the focus is on kinetic energy, then the description of the jet structure can be accomplished by two angular functions: the kinetic energy per unit solid angle\footnote{If the time $t$ at which the expression is valuated is such that the outflow is still in an acceleration phase, the appropriate Lorentz factor here is the ``terminal'' one, i.e., the one that can be estimated assuming the available internal (and/or magnetic) energy will be eventually converted to kinetic energy.}
\begin{equation}
 \frac{\mathrm{d}E}{\mathrm{d}\Omega}(\theta,t)=\int_0^\infty \left(\Gamma(r,\theta,t)-1\right)\Gamma(r,\theta,t)\rho'(r,\theta,t)c^2 r^2\mathrm{d}r 
 \label{eq:dE_dOmega}
\end{equation}
and the average Lorentz factor 
\begin{equation}
\Gamma(\theta,t)=\left(\frac{\mathrm{d}E}{\mathrm{d}\Omega}\right)^{-1}\int_0^\infty \left(\Gamma(r,\theta,t)-1\right)\Gamma^2(r,\theta,t)\rho'(r,\theta,t)c^2 r^2\mathrm{d}r,
\label{eq:Gammabulk}
\end{equation}
The latter description, applied to the ``coasting'' phase (see below), is the most widely adopted one, as it is sufficient for a basic description of the link between the prompt and afterglow emission observables and the jet structure in many contexts. The purpose of the above definitions is to clarify the connection between the three-dimensional physical properties of the outflow and the functions that are customarily used to describe its angular structure. A variety of similar, but not identical, definitions can be found in the literature: the essence of the arguments presented here does not depend on the precise definition.

\subsection{Stages in the life of a relativistic jet}\label{sec:stages}

In order to understand the structure of gamma-ray burst relativistic jets, we need to have a global view of how a jet is formed and how it evolves throughout its life. 
For that purpose, let us briefly summarize the main stages of the jet evolution with reference to the scheme shown in Fig.~\ref{fig:disegno}:
\begin{itemize}
    \item \textbf{Birth -- jet launch:} the jet is launched by the central engine. Different mechanisms have been considered to power the jet, depending on the nature of the central engine (e.g.\ an accreting BH \citep{Blandford1977-be} or a magnetar \citep{Usov1992Natur.357..472U}). Further details are provided in \S\ref{sec:central_engines};
    \item \textbf{Infancy -- jet head formation:} the jet material expands within the low-density funnel where it formed, until it collides with the dense ambient that surrounds the central engine (the progenitor star envelope or the merger ejecta), which we term `the progenitor vestige'. A forward-reverse shock structure forms -- the jet `head' \citep{Blandford1974} -- where the jet momentum flux is counterbalanced by the ram pressure of the vestige material (as seen in the head rest frame -- see \S\ref{sec:propagation});
    \item \textbf{Childhood -- jet propagation through the progenitor vestige:} the jet head, which is sustained by fresh jet material flowing across the reverse shock, propagates through the progenitor vestige \citep{Marti1994,Matzner2003MNRAS.345..575M,Morsony2007, Bromberg2011}. Due to the absence of lateral confinement, as soon as the head has slowed down enough as to become causally connected in the transverse direction, shocked material (both from the vestige and from the jet) is cast aside to form a hot, over-pressured cocoon that shrouds the jet and slowly expands laterally. The cocoon pressure is typically sufficient \citep{Bromberg2011} as to balance the lateral momentum flux of the jet material that flows from the central engine, leading to the formation of an oblique shock -- the `re-collimation' or `re-confinement' shock \citep{Falle1991MNRAS.250..581F} -- where the lateral component of the jet momentum is dissipated, turning the flow from radial into cylindrical. The jet is therefore collimated by its own cocoon (\S\ref{sec:propagation});
    \item \textbf{Adolescence -- breakout:} the jet head reaches the steep density gradient that marks the outer edge of the progenitor vestige. The head forward shock thus accelerates \citep{Shapiro1980,Tan2001}, the reverse shock disappears, and the jet and cocoon material start flowing freely out of the open channel: this process is broadly referred to as the `jet breakout'. During this process, the forward shock transitions from an optically thick region (where photon pressure dominates and the shock is therefore radiation-mediated) to an optically thin region: during this transition, photons from the hot downstream are released producing the `shock breakout' emission \cite{Gottlieb2018}, which represents the first observable electromagnetic emission in the jet's life. Childhood and adolescence are futher described in \S\ref{sec:propagation};
    \item \textbf{Adulthood -- free expansion}: the flow of fresh jet material from the central engine stops or diminishes significantly, setting a finite radial extent of the resulting outflow, which is now better described as an inhomogeneous shell \citep{Meszaros1993} that expands radially away from the progenitor at relativistic speed. After the jet breakout, the vast majority of the shell is still optically thick to Compton scattering \citep{Goodman1986,Baring1997,Lithwick2001-sc,Rees2005} (both off electrons associated to baryons in the outflow, and potentially off pairs that can form within the outflow as a consequence of energy dissipation events) for another few orders of magnitude in radius. Initially in this expansion phase radial density gradients remain frozen (`coasting phase', \citep{Meszaros1993,Kobayashi1999}) until radial pressure waves have the time to cross the outflow, leading to a radial spreading phase. During the free expansion phase, radial inhomogeneities in the bulk Lorentz factor can lead to the development of internal shocks \cite{Rees1994}, which have long been considered one of the main candidate mechanisms for the dissipation and subsequent radiation of the outflow's energy. Internal-shock-induced turbulence \cite{Zhang2011} has also been proposed as a possible triggering mechanism for magnetic reconnection (see also, e.g., \citep{Drenkhahn2002,McKinney2012}), which represents the other leading scenario for the dissipation of the outflow's energy in this phase; 
    \item \textbf{Seniority -- external shock:} the shell expands into the external low-density medium that surrounds the progenitor, which can be just the interstellar medium (ISM) or a stellar bubble inflated by the progenitor's stellar wind \citep{VanMarle2006}. As soon as the shell has swept a sufficient amount of external medium, corresponding to a rest mass energy equal to the shell's kinetic energy divided by the square of its bulk Lorentz factor \citep{Meszaros1993,Paczynski1993}, the expansion starts to be affected: a forward-reverse shock structure forms, with the reverse shock quickly crossing the entire shell \citep{Sari1995,Kobayashi2000_numerical}, initiating the deceleration of the latter and the transfer of its energy to the forward-shocked external medium. Soon after the start of the deceleration, the forward shock settles into a self-similar expansion phase \citep{Blandford1976,Kobayashi1999}, erasing any memory of the details of the shell radial structure. The angular structure remains unaffected as most of the shocked shell is out of causal contact in the transverse direction;
    \item \textbf{Senility -- lateral spreading and transition to non-relativistic expansion:} as soon as the transverse sound crossing time scale becomes shorter than the dynamical expansion time scale (or, in other terms, the angular size of causally connected regions starts to exceed the reciprocal of the local bulk Lorentz factor), pressure waves start to level out angular inhomogeneities, initiating a lateral expansion phase \citep{Paczynski1993,Rhoads1999,Kumar2003,Rossi2004,Zhang2009,VanEerten2010,Granot2012} which increases the shock working surface, therefore increasing the shock deceleration rate. The shock soon transitions to a non-relativistic expansion phase, slowly converging towards the Sedov-Taylor spherical blastwave behaviour.
\end{itemize}

\noindent The above brief account should help in making clear that the jet structure evolves throughout the life of a GRB jet. Three crucial phases can be identified in such an evolution: (1) the jet \textit{launch}, where the initial energy breakdown (internal, kinetic, magnetic), its angular profile and time dependence (all of which could contribute in principle to determine the properties of the jet at later stages) is set by the central engine and its evolution; (2) the \textit{interaction} of the jet with the progenitor vestige, comprising the \textit{propagation} of the jet head, the formation of the cocoon and the jet \textit{breakout}; (3) the \textit{expansion} of the jet in the external medium, and the subsequent \textit{deceleration}. The first phase sets the initial conditions of the problem and thus determines, along with the properties of the progenitor, the structure at later stages. On the other hand, it is possible that memory of the details of the initial conditions is lost along the evolution: as an example, as the jet transitions from adulthood to seniority (i.e.\ when the external shock enters the self-similar blastwave phase) the radial structure of the jet is erased (over a time scale that depends on the presence or absence of a low-velocity tail).

Given the time-dependent nature of the system, no univocal definition of jet structure exists: the most useful definition must be determined depending on the particular application and on the observational or theoretical aspects under consideration. Yet, in modelling the prompt and afterglow emission, often the most relevant structure is that corresponding to jet's adulthood, i.e.\ during the free expansion following the breakout from the progenitor vestige. This is the phase during which the prompt emission is believed to be produced, and the structure during this phase also constitutes the initial condition for producing the afterglow emission.  

\subsection{Models of jet-launching central engines}\label{sec:central_engines}
The leading jet-launching mechanism, by analogy with other relativistic jets such as those observed in active galactic nuclei (AGN, \cite{Blandford2018-vk}) and microquasars \citep{Molina_undated-de,Romero2016-uy}, is the Blandford-Znajek (BZ) mechanism \citep{Blandford1977-be}. This is a process by which the rotational energy of a spinning black hole (BH) is extracted in presence of a large-scale, poloidal magnetic field threading the horizon. The magnetic field is sustained by an accretion disk, and the mechanism requires the formation of a ``force-free'' \citep{Contopoulos2012-pv} magnetosphere close to the poles of the BH, which is only possible when the magnetic field energy density in the polar region exceeds the rest-mass energy density. In other words, a low-density funnel must be present along the BH rotation axis, and this is the region where the jet forms. The jet luminosity produced by the process follows \citep{Blandford1977-be}
\begin{equation}
 L_\mathrm{BZ} \propto B^2 M_\mathrm{BH}^2 a^2,
 \label{eq:BZpower}
\end{equation}
with higher-order ($a^4$) corrections when $a$ approaches unity \cite{Tchekhovskoy2010,Camilloni2022}. Here $B$ is the strength of the radial component of the magnetic field at the horizon, $M_\mathrm{BH}$ is the black hole gravitational mass and $a$ is its dimensionless spin parameter. The normalization constant depends on the magnetic field geometry and on the accretion disk properties \cite{Tchekhovskoy2010}. Typical expected values in the GRB context are $L_\mathrm{BZ}\sim 10^{49}-10^{51}$ erg/s, which match the loose observational luminosity constraints set by the observed gamma-ray energies and durations, and by the collimation angles inferred from afterglow observations (see \S\ref{sec:prompt_obs}). Jets launched by this process are expected to start off as magnetically-dominated outflows (i.e.\ $\sigma\gg 1$ at the jet base).  
 
Alternative jet-launching scenarios include energy deposition by neutrino - antineutrino ($\nu\bar\nu$) pair annihilation in the funnel above the BH \citep{Popham1999-zn, Ruffert1999,Chen2007-mt,Levinson2013-jv,Leng2014MNRAS.445L...1L,Salafia2021A&A...645A..93S}, and a proto-magnetar central engine \cite{Usov1992Natur.357..472U,Thompson2004ApJ...611..380T,Metzger2011MNRAS.413.2031M,Lu2015}. The $\nu\bar\nu$ luminosity for the former mechanism could be provided by the hot, inner parts of the accretion disk \cite{Zalamea2011MNRAS.410.2302Z}, and a jet powered by such process would be dominated by internal energy ($\eta/\rho'c^2\gg 1$) at its base, which would then be converted to kinetic energy by hydrodynamic acceleration. \respone{We note, though, that the $\nu\bar\nu$ mechanism seems unable to explain the large luminosities \citep{Kawanaka2013} and energies \citep{Leng2014} of some GRBs, and global simulations of jet launching in the aftermath of a binary neutron star merger seem to indicate that a jet powered by such mechanism would be unable to break out of the dense ejecta cloud that surrounds the merger remnant \citep{Just2016}.  }

\subsection{Models of jet propagation through the progenitor vestige}\label{sec:propagation}

\begin{figure}
 \centering
 \includegraphics[width=0.4\textwidth]{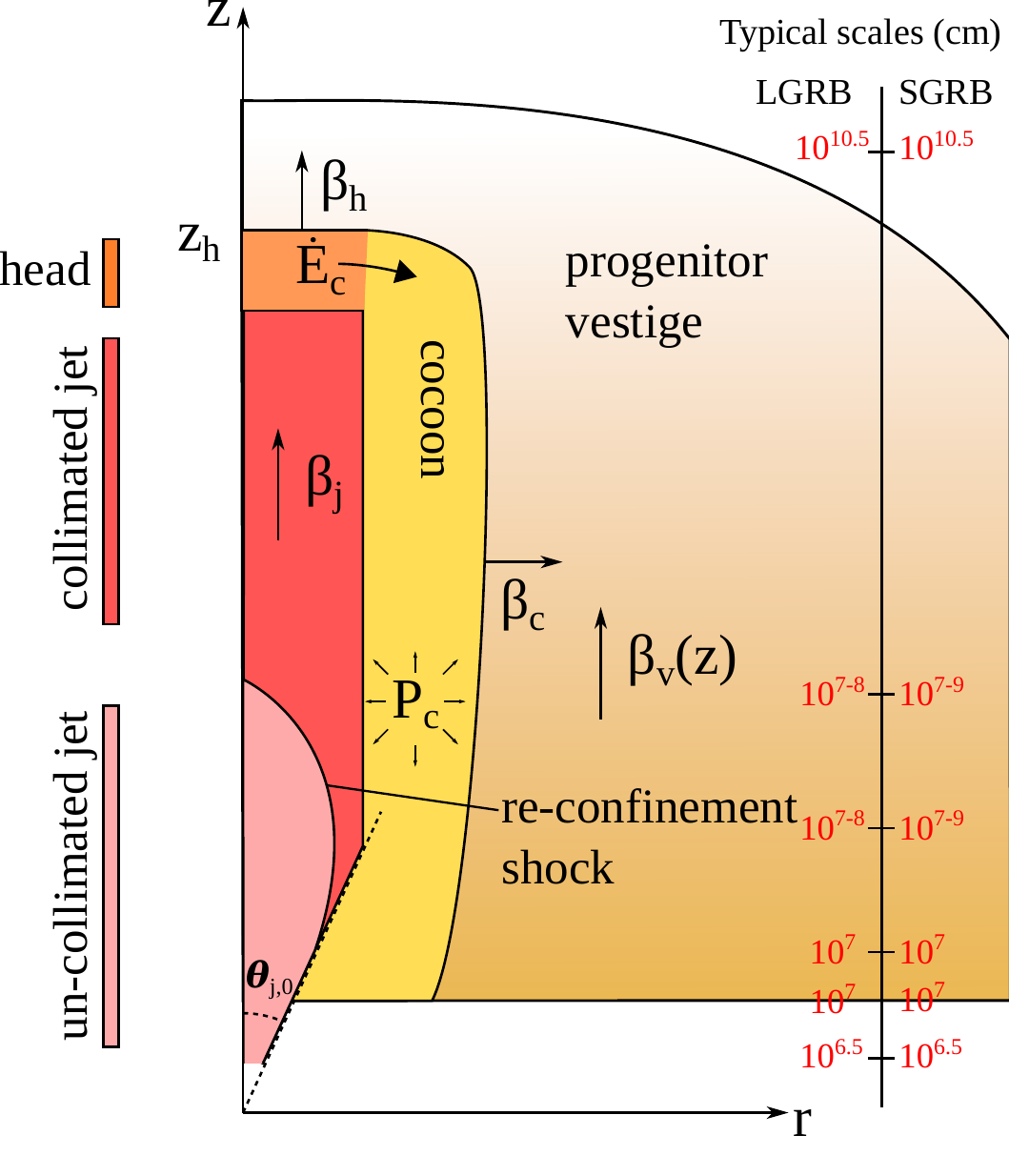}
 \caption{Sketch of the main elements in a basic hydrodynamical model of the jet propagation through the progenitor vestige. Adapted from \cite{Salafia2020_jet_propagation}.}
 \label{fig:jet_propag_sketch} 
\end{figure}

The jet launched by the central engine must initially propagate through the dense surrounding region constituted by the progenitor vestige, that is, the stellar envelope in the case of a collapsar or the ejecta cloud in the case of a compact binary merger. The details of the propagation and its final outcome depend on the properties of the jet at its base, on its duration, and on the properties of the vestige. The main features of the jet evolution during this phase can be unserstood based on a relatively simple hydrodynamical model (\cite{Falle1991MNRAS.250..581F,Morsony2007,Bromberg2014MNRAS.443.1532B,Salafia2020_jet_propagation,Hamidani2021MNRAS.500..627H,Gottlieb2021}, \respone{see e.g.\ \citep{Levinson2013,Bromberg2014,Bromberg2016} for the extension to the highly magnetized jet case, which presents some quantitative differences, despite the general picture remaining similar}), whose main features are the following (see Fig.~\ref{fig:jet_propag_sketch} for a sketch): the jet is represented by an outflow with luminosity $L_\mathrm{j}$, expanding radially at a speed $\beta_\mathrm{j,0}$ within an aperture angle $\theta_\mathrm{j,0}$ at its base. The outflow interacts with the vestige at a height $z$ above the central engine, where a forward-reverse shock structure forms, called the jet head. The jet head advancement speed $\beta_\mathrm{h}$ through the vestige, whose density at a height $z$ is $\rho_\mathrm{v}$ (and which can be expanding outwards at a speed $\beta_\mathrm{v}$), is set by the balance between the jet momentum flux that crosses the reverse shock and the ram pressure of the vestige material as seen from the forward shock downstram frame, namely (e.g. \cite{Begelman1989,Matzner2003MNRAS.345..575M,Bromberg2011,Murguia2017ApJ...835L..34M}, neglecting the vestige pressure)
\begin{equation}
 \Gamma_\mathrm{j}^2\Gamma_\mathrm{h}^2(\beta_\mathrm{j}-\beta_\mathrm{h})^2\rho'_\mathrm{j}h_\mathrm{j}c^2 = \Gamma_\mathrm{h}^2\Gamma_\mathrm{v}^2(\beta_\mathrm{h}-\beta_\mathrm{v})^2\rho_\mathrm{v}c^2,
\end{equation}
where $\Gamma_\mathrm{x} = (1-\beta_\mathrm{x}^2)^{-1/2}$, $\beta_\mathrm{j}$ is the  dimensionless jet speed just before crossing the reverse shock, $\pi\theta_\mathrm{j}^2z^2$ is the reverse shock working surface, and we used $\rho'_\mathrm{j}h_\mathrm{j}=L_\mathrm{j}/\pi\theta_\mathrm{j}^2z^2\beta_\mathrm{j}\Gamma_\mathrm{j}^2 c^3$. This can be solved for $\beta_\mathrm{h}$, which gives
\begin{equation}
 \beta_\mathrm{h}=\frac{\beta_\mathrm{j} + \tilde L^{-1/2}\beta_\mathrm{v}}{1+\tilde L^{-1/2}},
\end{equation}
where the dimensionless quantity $\tilde L = \Gamma_\mathrm{j}^2\rho'_\mathrm{j}h_j/\Gamma^2_\mathrm{v}\rho_\mathrm{v}=L_\mathrm{j}/\pi\theta_\mathrm{j}^2z^2\beta_\mathrm{j}\Gamma_\mathrm{v}^2\rho_\mathrm{v} c^3$ is what sets the overall properties of the jet avancement \cite{Bromberg2014MNRAS.443.1532B}. When the head advancement is sub-relativisic, then $\beta_\mathrm{h}\sim \beta_\mathrm{v}+\tilde L^{1/2}\beta_\mathrm{j}$; when it is relativisic, then $\Gamma_\mathrm{h}\sim \tilde L^{1/4}/\sqrt{2}$.

If the head advancement speed $\beta_\mathrm{h}$ is sufficiently low, $\Gamma_\mathrm{h}\beta_\mathrm{h}\lesssim 1/\sqrt{3}\theta_\mathrm{j}$ \cite{Salafia2020_jet_propagation}, the head is causally connected by sound waves in the transverse direction: the lack of lateral confinement in the head then causes the shocked material (both that of the jet and that of the vestige) to flow laterally and form an over-pressured cocoon that shrouds the jet. The cocoon slowly expands laterally within the vestige at a speed proportional to the square root of the ratio of its pressure to the average vestige density \cite{Begelman1989,Bromberg2011}, but this is typically not sufficient to prevent a pressure build-up as it is filled with an increasing amount of shocked material flowing from the head. When the cocoon pressure becomes comparable to the transverse momentum flux of the jet at its base, a ``re-confinement shock'' forms \cite{Falle1991MNRAS.250..581F} where such transverse momentum is dissipated, collimating the jet into a cylindrical flow. The condition for such a self-collimation can be written approximately as $\tilde L \lesssim \theta_\mathrm{j,0}^{-4/3}$ \cite{Bromberg2011}. The self-collimation reduces the reverse shock working surface, therefore favouring the jet head propagation. The increase in the head speed, on the other hand, reduces the energy flow to the cocoon, therefore affecting its ability to effectively collimate the jet. In self-collimated jets, the final jet opening angle at breakout is thus set by these competing effects. 

In presence of a homologous expansion of the vestige (as expected in the case of compact binary merger ejecta), another self-regulation effect arises \cite{Duffell2018ApJ...866....3D}: if the jet head stalls (i.e.\ $\beta_\mathrm{h}\sim \beta_\mathrm{v}$) and the jet is self-collimated (so that the head working surface is constant -- but this is ensured by the fact that $\beta_\mathrm{h}\sim\beta_\mathrm{v}$ implies $\tilde L\ll 1$ and hence the self-collimation condition $\tilde L\lesssim \theta_\mathrm{j,0}^{-4/3}$ is certainly satisfied), the expansion has the effect of easing the head propagation, because it reduces $\tilde L$. If the jet is launched shortly after the onset of the vestige homologous expansion (so that the jet duration is much longer than such delay), the result is that the jet's ability to break out depends solely on the ratio of the jet energy to that of the expanding vestige \cite{Duffell2018ApJ...866....3D}, regardless of the jet duration.

The importance of the jet propagation phase, from the observational point of view, stems from the fact that the jet structure after breakout carries the imprint of the jet-vestige interaction. If the rearrangement of the jet after breakout (and before the prompt emission is produced) does not erase such memory, it is possible in principle to extract information about the progenitor (and possibly also about the central engine) from the jet structure at a stage when observable emission can be produced (e.g. \cite{Matzner2003MNRAS.345..575M}). The extent to which the resulting structure is determined by the central engine, the jet-vestige interaction, or both, is still a matter of debate. For what concerns binary neutron star mergers, the recent three-dimensional, special-relativistic hydrodynamical numerical simulations by \cite{Nativi2022MNRAS.509..903N} seem to suggest that turbulence at the jet-cocoon interface (see also \cite{Gottlieb2021-ro}) tend to erase the details of the injected jet structure (i.e.\ the angular structure as initially set by the central engine), which could lend support to the hypothesis that ``adult'' jets from binary neutron star mergers share a quasi-universal structure (e.g. \cite{Rossi2002,Salafia2015-dj}). Yet, the development of such turbulence seems strongly suppressed in magnetohdrodynamic simulations with a significantly magnetized injected jet \cite{Gottlieb2021MNRAS.504.3947G} and, moreover, simulations that use a different jet injection technique (e.g. \cite{Urrutia2021MNRAS.503.4363U}) come to the opposite conclusion. 

\subsection{Jet and cocoon breakout}\label{sec:breakout}
When the jet head forward shock reaches the steep density gradient that characterizes the outer edge of the progenitor vestige, it starts accelerating in much the same way as a supernova shock does as it approaches the outer edge of the stellar envelope \cite{Gandelman1956,Colgate1974,Falk1978,Shapiro1980,Matzner1999}. The main differences between the jet forward shock breakout and a supernova shock breakout are that the former is relativistic and highly anisotropic, both features having a strong impact on the resulting dynamics and emission \cite{Tan2001,Nakar2012,Matzner2013,Yalinewich2017,Linial2019,Irwin2021}. As explained in \S\ref{sec:stages}, the head forward shock separates the shocked vestige material from the unshocked one (see Fig.~\ref{fig:disegno}). Below the inner part of the forward shock, within an angle $\theta_\mathrm{j,bo}$ (the \textit{jet angle at breakout}), is the head reverse shock: since the jet material crossing the latter is (by definition) faster than the head, its ram pressure ensures that the reverse shock keeps up with the accelerating forward shock. As a result, the head material remains dense and optically thick to Compton scattering during the breakout, and therefore its internal energy contributes to the expansion rather than being radiated. At angles larger than $\theta_\mathrm{j,bo}$, on the other hand, the forward shock breakout is accompanied by the expansion of the underlying shocked material (which is part of the upper cocoon): the forward shock transitions from radiation-mediated to collisionless, liberating photons in what is sometimes called the ``cocoon shock breakout emission'' \cite{Gottlieb2018,Gottlieb2018-vx,Nakar2017} (with a typical temperature of several tens of keV, \citep{Nakar2012}), and the underlying material becomes gradually transparent, giving rise to a ``cooling'' emission (typically peaking in the UV, \cite{Nakar2012}).  The expansion of the cocoon after the shock breakout is sometimes called the cocoon ``blowout'' \cite{Matzner2003MNRAS.345..575M,Salafia2020_jet_propagation}. The entire breakout process is followed by a rearrangement of the jet and cocoon material into an inhomogeneous shell, which is what is most often referred to as the structured jet. In the following section, we focus on the structural properties of the latter, as found from numerical simulations.

\subsection{Expected general features of the jet structure in GRBs}

\begin{figure}
 \centering
\includegraphics[width=\columnwidth]{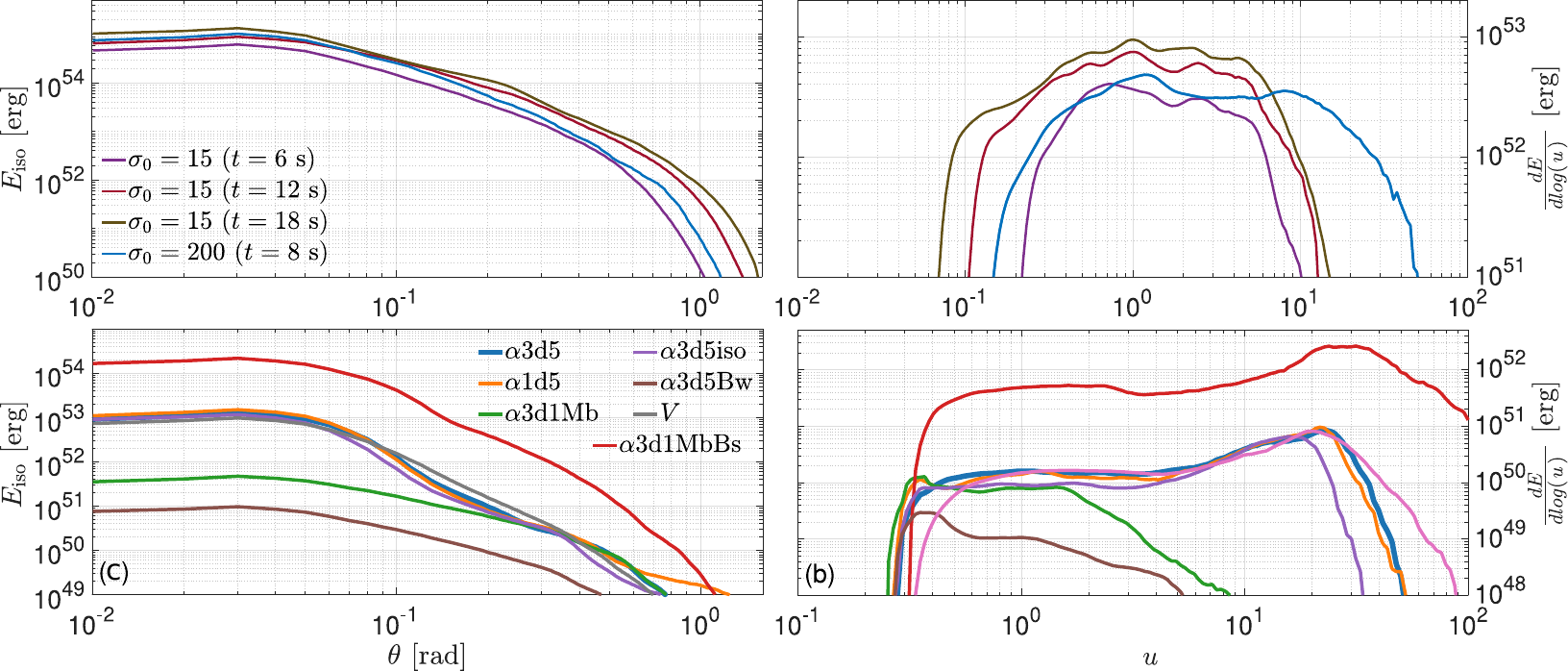}\caption{Jet structures resulting from GRMHD simulations representing collapsars (upper panels, adapted from \cite{Gottlieb2022_LGRB_GRMHD}) and binary neutron star merger remnants (lower panel, adapted from \cite{Gottlieb2022_SGRB_GRMHD}). Left-hand panels show the isotropic-equivalent energy $E_\mathrm{iso}=4\pi\mathrm{d}E/\mathrm{d}\Omega$ from a late snapshot of the simulations (significantly later than the jet breakout), while right-hand panels show the distribution of energy in the four-velocity modulus $u=\Gamma\beta$ space. Different colors refer to different initial conditions, as detailed in \cite{Gottlieb2022_LGRB_GRMHD} and \cite{Gottlieb2022_SGRB_GRMHD}.}
\label{fig:Gottlieb_GRMHD}
\end{figure}

Despite the complexity of the involved processes, the modelling and understanding of the birth and evolution of relativistic jets has been addressed since the late seventies, initially prompted by observations of radio galaxies \cite{Blandford1974}. A big deal of the understanding of the jet launching and its evolution has been obtained through numerical simulations of the central engine (typically within general-relativistic magnetohdrodynamics, GRMHD) and of the jet propagation (usually within a special-relativisic hydrodynamics -- RHD -- or magnetohdrodynamics -- RMHD -- framework, see \cite{Komissarov2021} for a recent review). In the GRB context, a widely adopted approach (due in part to computational limitations) to the simulation of the propagation and breakout phase has been that of ``injecting'' a jet with properties based on an educated guess into a model of the progenitor vestige (in two dimensions, e.g. \cite{Zhang2003,Morsony2007}, and three dimensions, e.g.\ \cite{LopezCamara2014,Gottlieb2020_struct_magjets,Gottlieb2021,Nativi2022MNRAS.509..903N}, but see e.g.\ \cite{Aloy2005,Just2016,Bromberg2016}). While this helps in limiting the needed computational resources by leaving out the central engine region from the computational domain, it prevents a direct connection between the properties and evolution of the central engine and those of the jet at larger scales: in particular, this approach does not allow for a self-consistent description of (1) the central engine variability (due e.g.\ to the stochastic fluctuations in the accretion rate due to a turbulent disk), (2) the evolution of the jet luminosity (linked to that of the accretion rate and/or of the central compact object) and possibly (3) orientation,  the (4) injected jet structure (angular distributions of kinetic luminosity and magnetization) and (5) the effects of the central engine on the vestige (e.g.\ gravity and the accretion disk winds). On top of this, the idealized nature of the progenitor models employed in most of these studies can affect the results by introducing exact symmetries that are not present in nature. The steady advancement of computational methods and resources has led recently to many important works that investigated some of these limitations (e.g.\ \cite{Kathirgamaraju2018,Kathirgamaraju2019,Fernandez2019,Christie2019, Ito2021,Pavan2021,Nathanail2021,Gottlieb2022_LGRB_GRMHD,Gottlieb2022_SGRB_GRMHD}). These include three-dimensional GRMHD simulations that self-consistently cover jet launch (usually within the Blandford-Znajek paradigm), propagation and breakout \cite{Nathanail2021,Gottlieb2022_LGRB_GRMHD,Gottlieb2022_SGRB_GRMHD}, even though these still feature idealized initial conditions and do not include a treatment of neutrinos, whose contribution to cooling, transport of momentum and energy can have prominent effects on the central regions. Yet, these simulations currently constitute some of the most detailed investigations that can shed light on the GRB jet structure. Fig.~\ref{fig:Gottlieb_GRMHD} shows the jet structures obtained from GRMHD simulations in the collapsar case (top panels - \cite{Gottlieb2022_LGRB_GRMHD}) and in the case of binary neutron star mergers (bottom panels - \cite{Gottlieb2022_SGRB_GRMHD}). In qualitative agreement with previous works, these investigations find a jet angular structure after breakout that broadly features a narrow \textit{core} (with an opening angle of few degrees) with approximately uniform Lorentz factor and energy density (mostly containing jet material that crossed the collimation shock, but did not reach the head before breakout), surrounded by a wider structure (sometimes called the jet \textit{wings}, and typically composed of an inner part of mixed jet and cocoon material -- where the amount of mixing depends on the jet magnetization, e.g.\ \cite{Gottlieb2020_struct_magjets} -- surrounded by a wider, cocoon-dominated part) where both the average Lorentz factor and the energy density fall off relatively quickly with the angle  (typically as steep power laws $\propto\theta^a$ with $a\lesssim -3$, or as Gaussians).  A minority of the simulations find a ``hollow'' jet core with a lower energy density and Lorentz factor along the axis with respect to that at the core edge \cite{Nathanail2021}, but it is unclear whether this is a genuine result or a numerical artifact \cite{Zhang2004}. 

\section{Prompt emission from a structured jet}
The origin of the prompt emission of GRBs is not well understood yet. Generally speaking, the emission is thought to be powered by some mechanism that dissipates the dominant form of energy in the jet (either kinetic or magnetic), transforming it into internal energy. In most scenarios, part of the latter takes the form of relativistic particles with a non-thermal energy distribution, which produce the observed emission by means of some radiative process at the photosphere or beyond \cite{Bosnjak2022Galax..10...38B}. Different dissipation mechanisms operating within the relativistic outflow and different radiation mechanisms (typically synchotron or inverse Compton) have been proposed, but no compelling evidence has been found yet in support of any of the envisaged scenarios. Still, the jet structure might have a leading role in explaining at least some of the prompt emission features. 

\subsection{Observed temporal and spectral poperties of GRB prompt emission}\label{sec:prompt_obs}
The GRB observed duration distribution is bimodal, which leads to the division into short and long \emph{duration} events (with an observer frame separation at $\sim$2 sec). The average spectral properties of events in the two duration classes display some differences, with short GRBs featuring on average harder spectra with respect to long ones \cite{Kouveliotou1993}, the difference residing primarily in the low-energy part of the spectrum \cite{Ghirlanda2004-qm,Ghirlanda2011-zf}. Short events are detected at a lower rate with respect to long events (with an instrument-dependent short/long ratio of 1:3 for \textit{CGRO}/BATSE, 1:5 for \textit{Fermi}/GBM and of 1:9 for \textit{Swift}/BAT). However, several different instrumental and cosmological effects shape (and hence bias) these observed properties \cite{Sakamoto2008,Qin2012-en,Ghirlanda2015}.  

The prompt emission has no apparent periodicity \cite{Guidorzi2016-pw} and features a power density spectrum consistent \cite{Beloborodov2000-pi,Guidorzi2012-eu} with turbulent dissipation processes. Different methods were employed to measure the minimum \emph{variability} timescale of prompt emission light curves, resulting in different distributions \cite{MacLachlan2013-jc,Zach_Golkhou2014-pk}. Ref.~\cite{Zach_Golkhou2014-pk} reports similar rest-frame distributions for the minimum variability timescale in long and short GRBs, centered around 0.5 s and extending down to 10 ms in $\sim$10\% of the events. Another interesting feature of the prompt emission is the presence of \emph{spectral lags}. These consist in the fact that pulses observed in the lower energy bands of the gamma-ray instruments are seen to lag behind the corresponding pulses in the higher energy bands \cite{Norris2005-zt}. This feature seems to be more commonly present in long GRBs, while short events typically have lags consistent with zero \cite{Bernardini2015-jx}. 

Most attempts at identifying the fundamental building blocks of GRB prompt emission light curves adopted parametric functions to represent pulses (e.g. \cite{Norris2005-zt,Kocevski2003-xq}). With the caveat that these methods are applied to large samples of light curves of GRBs with unknown redshift, the results show apparent (observer frame) differences between short and long GRBs \cite{Bhat2011-ry}. 

On the longer timescales, periods of activity can at times be separated by quiescent phases. In a sizable fraction ($\sim 15\%$) of long GRBs, a long quiescence phase (reaching, in some cases, $>$100 sec) separates \emph{precursor} activity from the main emission episode \cite{Burlon2008-fe}. Precursors have been identified also in short GRBs  \cite{Troja2010-sj}. Long apparent quiescences separate also the main event from late time pulses, or \emph{flares}, often observed in the X-ray band by \textit{Swift}/XRT. X-ray flares share some common properties with the prompt emission \cite{Pescalli2018-nc} and are thus often interpreted as linked to late-time central engine activity. No significant differences in the average spectral properties of precursors and main emission episodes have been identified \cite{Burlon2009-pk}, while flares appear clearly softer. 

The prompt emission of GRBs is characterized by a \emph{non-thermal} spectrum. Presence of thermal-like emission has been identified in few cases either during the initial phases of the burst \cite{Ghirlanda2003-xp} or along its full duration \cite{Ghirlanda2013-wy}, with no evidence of such emission component in short bursts \cite{Lazzati2005-ar}. The combination of multiple emission components (e.g.\ the sum of a powerlaw with a high energy cutoff and a black body) were also adopted to interpret observed spectra \cite{Ryde2009-kg,Guiriec2015-hb}. Observationally, the spectral energy distribution of GRB prompt emission typically peaks at 0.1-1 MeV and the low (resp.\ high) energy spectrum, below (resp.\ above) the peak, is consistent with a power-law with photon index $\sim$-1 (resp.\ -2.5).  

\subsection{On the synchrotron origin of GRB prompt emission}
The interpretation of the prompt emission spectrum as synchrotron from shock-accelerated electrons faced the contradicting evidence of the observed GRB spectra being harder, below the SED peak, than the expected synchrotron spectral shape. Given the physical conditions of the emission region, in particular a large magnetic field $B\sim10^{4-6}$ Gauss, shock accelerated electrons should cool rapidly \cite{Ghisellini2000-eg} producing a spectrum with photon index $-1.5$ below the synchrotron characteristic SED peak. 
One possible solution to this issue \cite{Derishev2001A&A...372.1071D,Nakar2009ApJ...703..675N,Daigne2011A&A...526A.110D} considered that electrons do not cool efficiently (so-called marginally fast cooling scenario) so that the separation between the characteristic synchrotron frequency (identified as the peak of the $\nu F_{\nu}$ spectrum) and the cooling frequency is relatively small. As such, the hardest synchrotron spectral powerlaw (i.e. the single electron spectrum with photon index -2/3) would become visible in the observer energy range. The tension with observations would be solved admitting that the fitted empirical Band function captures an average spectral index between the two characteristic ones (i.e. -2/3 and -3/2 below and above the cooling break respectively) \cite{Toffano2021-rv}. This interpretation was recently proved valid by the discovery  \cite{Oganesyan2018-oy,Oganesyan2019-zl,Ravasio2018-gn,Ravasio2019-aa} in long {\it Swift} and {\it Fermi} GRBs of a spectral break distributed in the 1-100 keV range. Overall, these studies find that in a sizable fraction of bright GRBs (possibly limited by current detectors' performances - see \citep{Toffano2021-rv}) there is a break, located at energies a factor $\sim$10 below the characteristic SED peak. Remarkably, the power-law indices below and above the break are consistent with the single electron synchrotron photon spectral index (-2/3) and cooling synchrotron photon index (-3/2) respectively. Further support to the synchrotron origin of the prompt emission was obtained by fitting a  synchrotron model directly to the data \cite{Ronchi2020-ex,Michael_Burgess2019-lq}.

\begin{figure}
    \centering
    \includegraphics{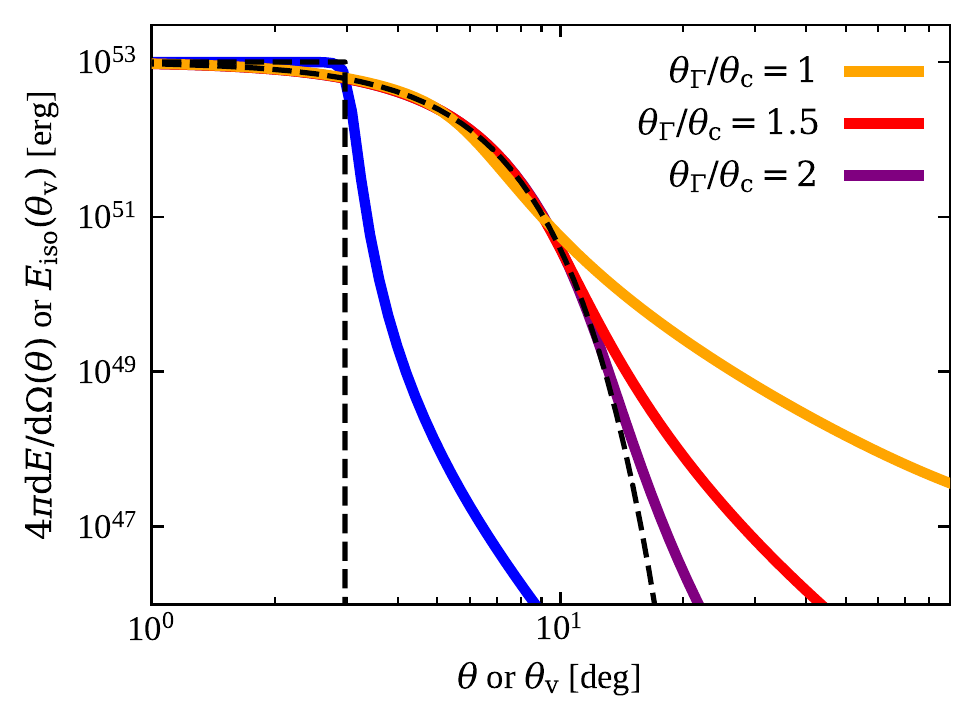}
    \caption{Apparent and intrinsic structure for a uniform and a Gaussian structured jet. Black dashed lines represent $4\pi \mathrm{d}E/\mathrm{d}\Omega$ at an angle $\theta=\theta_\mathrm{v}$ from the jet axis. Colored solid lines show $E_\mathrm{iso}(\theta_\mathrm{v})$ for four different models: a uniform jet model with $\mathrm{d}E/\mathrm{d}\Omega = 10^{53}/4\pi$ erg and $\Gamma = 300$  within an angle $\theta_\mathrm{j}=3^\circ$ (blue) and three Gaussian models with $\mathrm{d}E/\mathrm{d}\Omega = (10^{53}/4\pi) \exp[-(\theta/\theta_\mathrm{c})^2/2]$ and $\Gamma\beta = 300\exp[-(\theta/\theta_\Gamma)^2/2]$, with $\theta_\mathrm{c}=3^\circ$ and three different values of $\theta_\Gamma$ (orange, red and purple, with the corresponding $\theta_\Gamma/\theta_\mathrm{c}$ ratios given in the legend). Adapted from \cite{Salafia2015-dj}.}
    \label{fig:intrinsic-apparent}
\end{figure}

While these results, after three decades of debate, represent a step forward to unveil the synchrotron nature of the prompt emission, they present further challenges \citep{Ghisellini2020-ok}. If the break is interpreted as the cooling synchrotron frequency, it implies a small magnetic field ($B\sim 10$ G) in the emission region. If the latter is located relatively close to the central engine (as suggested by the observed small variability timescales) the Synchrotron Self Compton (SSC) emission would become relevant though its signature has not been clearly observed at GeV energies by {\it Fermi}/LAT. Possible solutions, consider emission in a downstream decaying magnetic field (e.g. \citep{Uhm2014NatPh..10..351U}) or proton-synchrotron emission \cite{Ghisellini2020-ok} (but see \citep{Florou2021-nh}).

\subsection{Correlations between spectral peak frequency and energetics}
Other key features of the prompt emission, common to both long and short GRBs, are the observed correlations between the rest frame SED peak energy ($E_{\rm peak}$) and the burst energy or luminosity, considering isotropic emission ($E_{\rm iso}$ - \citep{Amati2002-qi} and $L_{\rm iso}$ \citep{Yonetoku2004-zv}) or accounting for the GRB jet aperture angle (i.e.\ $E_{\gamma}=E_{\rm iso}(1-\cos\theta_\mathrm{j})$, see \citep{Ghirlanda2004-ny}). Short and long GRBs, owing to their different duration, follow two nearly parallel correlations in the $E_{\rm peak}-E_{\rm iso}$ plane  while they show a similar $E_{\rm peak}-L_{\rm iso}$ correlation \cite{DAvanzo2014-mo,Ghirlanda2015-mp}. While these correlations may be subject to instrumental selection effects, their physical nature is corroborated by the existence, within individual GRBs, of similar relations between the same observables as a function of time along the prompt emission duration 
\cite{Ghirlanda2011-zf,Calderone2015-jm}.

\subsection{Impact of jet structure on the prompt emission observables}

One big difficulty in interpreting the prompt emission properties of GRBs in terms of the jet structure  comes from the fact that the prompt emission mechanism is not well understood: for this reason, a typical approach is to assume that the prompt emission simply transforms some fraction of the kinetic (or magnetic) energy into radiation. Given a prompt emission efficiency $\eta_\gamma(\theta)$ (which represents the fraction of the available kinetic/magnetic energy at $\theta$ that is radiated in gamma-rays), the dependence of the prompt emission properties ($E_{\rm peak}$, $E_{\rm iso}$, $L_{\rm iso}$) on the viewing angle $\theta_\mathrm{v}$ (the \textit{apparent structure} in the language of \cite{Salafia2015-dj,Salafia2015-xx}) is set by the energy and Lorentz factor angular profiles, Eqs.~\ref{eq:dE_dOmega} and \ref{eq:Gammabulk}. In particular, the bolometric isotropic-equivalent energy can be computed as \cite{Salafia2015-dj,Beniamini2019-no,Ioka2019}
\begin{equation}
 E_\mathrm{iso}(\theta_\mathrm{v}) = \int_0^{2\pi}\mathrm{d}\phi\int_{0}^{\pi/2}\sin\theta\,\mathrm{d}\theta\,\frac{\delta^3(\theta,\phi,\theta_\mathrm{v})}{\Gamma(\theta)}\eta_\gamma(\theta)\frac{\mathrm{d}E}{\mathrm{d}\Omega}(\theta),
 \label{eq:Eiso_thv}
\end{equation}
where $\delta=\Gamma^{-1}(1-\beta\cos\alpha)^{-1}$ is the Doppler factor, with $\alpha$ being the angle between the line of sight and the radial expansion direction, which can be expressed as \cite{Salafia2015-dj}
\begin{equation}
 \cos\alpha = \cos\theta\cos\theta_\mathrm{v} + \sin\theta\sin\phi\sin\theta_\mathrm{v}.
\end{equation}

Figure \ref{fig:intrinsic-apparent} shows $E_\mathrm{iso}(\theta_\mathrm{v})$ corresponding to a few different $\mathrm{d}E/\mathrm{d}\Omega$ and $\Gamma$ profiles, assuming a constant $\eta_\gamma$ at all angles. A general feature that is demonstrated in the figure is that at some viewing angles (typically close to the jet axis) the emission is dominated by material moving along the line of sight, resulting in  $E_\mathrm{iso}(\theta_\mathrm{v})\sim 4\pi \mathrm{d}E/\mathrm{d}\Omega(\theta=\theta_\mathrm{v})$. Far off-axis, on the other hand, the flux received by the observer is spread over a larger portion of the jet, corresponding to regions with the most favourable combination of a large intrinsic luminosity and a sufficiently low Lorentz factor as to avoid a too severe de-beaming of radiation away from the line of sight. The steeper the Lorentz factor decay as a function of $\theta$, the shallower the $E_\mathrm{iso}$ decay at large $\theta_\mathrm{v}$, as the de-beaming is less severe in broader and more energetic portions of the jet.

In order to compute $E_\mathrm{peak}(\theta_\mathrm{v})$, one needs to make a further assumption about the spectral shape $S'_\nu(\nu',\theta)$ of the radiation as measured in the jet comoving frame (e.g.\ \cite{Salafia2019_onaxisview}, \cite{Ioka2019}): the simplest assumption, often adopted in the literature, is that of a fixed spectrum at all angles, which yields $E_\mathrm{peak}(\theta_\mathrm{v})\propto \Gamma(\theta=\theta_\mathrm{v})$ for viewing angles at which the emission is dominated by material on the line of sight, and a shallower decrease at larger viewing angles (see e.g.\ Figure 2 in \cite{Salafia2019_onaxisview}). In long GRBs, this simple assumption (along with a Gaussian ansatz for the jet structure) is sufficient \cite{Salafia2015-dj} to reproduce the observed $E_\mathrm{iso}-E_\mathrm{peak}$ correlation \cite{Amati2002-qi}, while in short GRBs it leads to a clear tension with the observed $E_\mathrm{peak}$ of GRB170817A \cite{Ioka2019}. A more `physical' approach to the modelling of the expected peak spectral energy would require one to assume a particular prompt emission scenario, such as e.g.\ that of internal shocks, and explicitly compute the typical comoving photon energy at each angle for a given jet structure \cite{Zhang2002Epeak}.

The computation of $L_\mathrm{iso}(\theta_\mathrm{v})$ also requires additional assumptions \citep{Kathirgamaraju2018,Kathirgamaraju2019}. The duration of the prompt emission is usually assumed to be linked to that of the central engine activity. This is based on the idea that the prompt emission is composed of short pulses, each corresponding to a dissipation event in the jet. The events happen around a typical radius $R$, where the jet material travels with a typical bulk Lorentz factor $\Gamma$, and their intrinsic duration is very short, so that the observed duration is set by the (Doppler-contracted) angular time scale $t_\mathrm{ang}\sim R/\Gamma^2 c$ (i.e.\ the maximal arrival time difference between photons emitted within an angle $1/\Gamma$ with respect to the line of sight), which is assumed to be much shorter than the central engine activity duration $T_\mathrm{CE}$. As a result, the observed duration is $T_\mathrm{GRB}\sim T_\mathrm{CE}$ and the average isotropic-equivalent luminosity is simply $L_\mathrm{iso}\sim E_\mathrm{iso}/T_\mathrm{CE}$. If the spread in $R$ and $\Gamma$ is not too large, then this holds true also for off-axis observers, as long as the single-pulse duration remains much shorter than $T_\mathrm{CE}$ \cite{Salafia2016}, and in such situation Eq.~\ref{eq:Eiso_thv} divided by $T_\mathrm{CE}$ can be used to compute the average $L_\mathrm{iso}(\theta_\mathrm{v})$ \cite{Barbieri2019,Ioka2019,Salafia2019_onaxisview,Salafia2020_jet_propagation,Patricelli2022}. Pulse overlap can result in a shallower dependence on the viewing angle \cite{Salafia2016}, but the dependence will eventually steepen at sufficiently large angles such that the duration of single pulses will become comparable or longer to $T_\mathrm{CE}$.

Useful discussions on the transformation between on- and off-axis isotropic-equivalent energies, durations, luminosities and peak photon energies, along with useful analytical approximations, can be found in \cite{Matsumoto2019a,Matsumoto2019b}. 

\subsection{The GRB luminosity function and the jet structure}

\begin{figure}
    \centering
    \includegraphics[width=10.8cm]{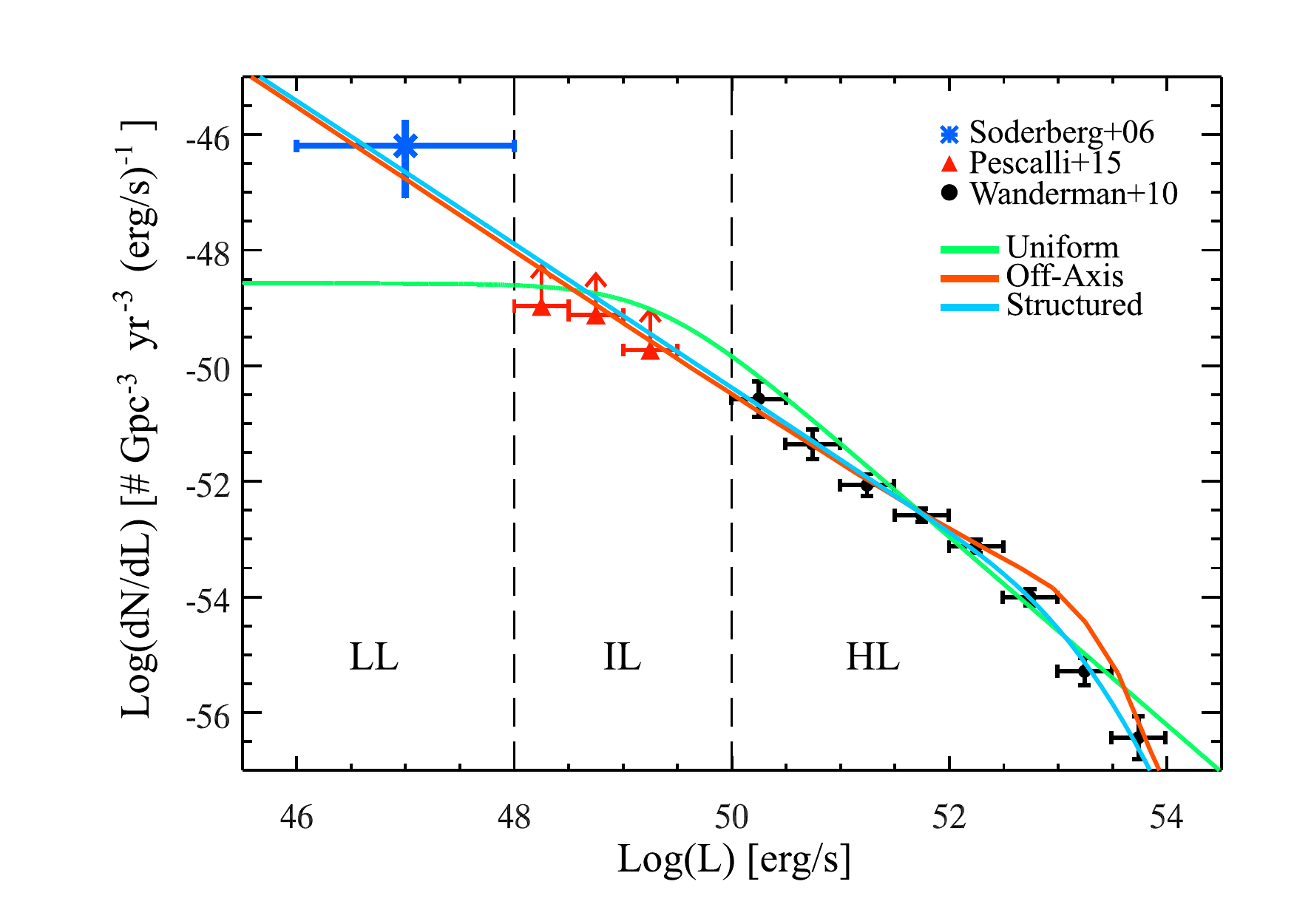}
    \includegraphics[width=10cm]{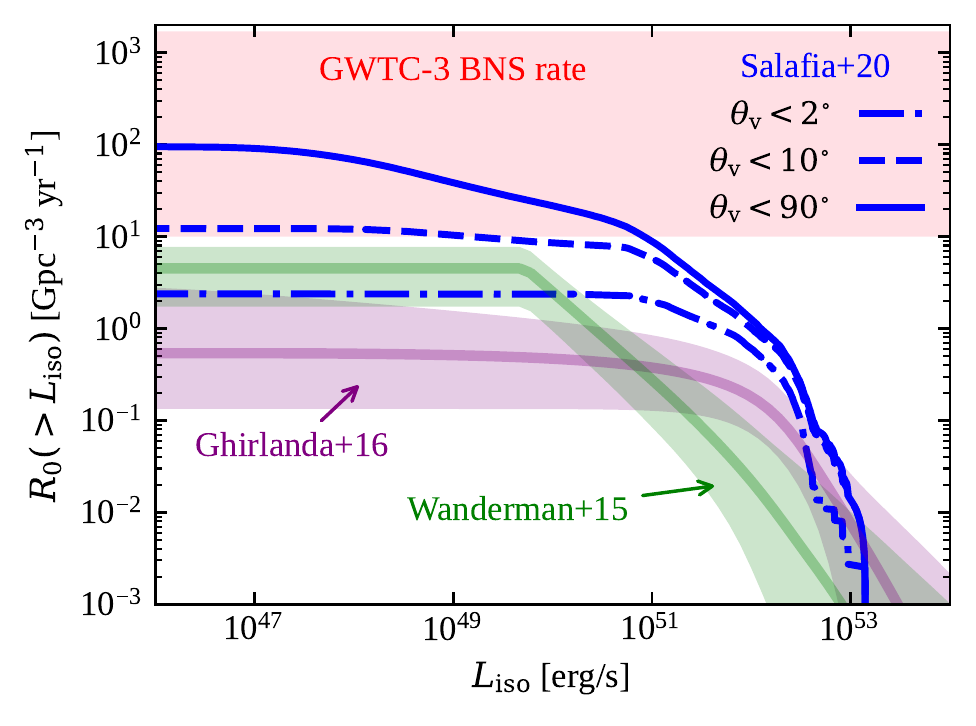}
    \caption{{\it Top:} luminosity function of Long GRBs as obtained by \cite{Wanderman2010-cb} (black symbols) extended to low luminosities by \cite{Pescalli2015-bm} (red and blue sumbols). Models considering a uniform jet (only seen on-axis -- green -- or isotropically oriented  -- red) or a structured jet with a steep power law profile (cyan) are shown. The separation in low, intermediate, high luminosity (LL, IL, HL) GRBs is indicated by the dashed vertical lines. {\it Bottom:} models of the (inverse cumulative) SGRB luminosity function. Models fitted to observed properties of short GRBs (detected at cosmological distances) are shown by the green \cite{Wanderman2015-ji} and purple \cite{Ghirlanda2016-hm} thick transparent lines and bands (medians and 90\% credible regions). The luminosity function obtained by \cite{Salafia2020_jet_propagation} by computing the jet structure from a semi-analitical calculation of the jet propagation and breakout is shown by the blue  lines (contributions by jets observed in different intervals of viewing angle are shown), arbitrarily normalized to a local rate density $R_0=100\,\mathrm{Gpc^{-3}\,yr^{-1}}$. The local BNS merger rate density constraint from \cite{LVC2021_gwtc3_pop}, i.e.\ $10\leq R_\mathrm{0,BNS}/\mathrm{Gpc^3\,yr}\leq 1700$, is shown by the pink shaded region. 
    }
    \label{fig:LF}
\end{figure} 

A key property of the GRB population is the luminosity function (LF). The LF can be defined as the probability density of the isotropic-equivalent luminosity at a particular redshift $P(L_\mathrm{iso}\,|\,z)$ or, equivalently, as the comoving rate density in a differential luminosity bin, $\mathrm{d}R_z/\mathrm{d}L_\mathrm{iso}=R_z P(L_\mathrm{iso}\,|\,z)$, where $R_z$ is the event rate density (GRBs per comoving Gpc$^3$ yr) at redshift $z$. If the population does not feature a luminosity evolution with redshift, then the local LF $\mathrm{d}R_0/\mathrm{d}L_\mathrm{iso} = R_0 P(L_\mathrm{iso})$ is sufficient as to describe the luminosity distribution in the population. In the context of a structured jet, the luminosity of each event depends both on the intrinsic properties of the underlying jet, and on its viewing angle. Hence, the lumnosity function is shaped at least in part by viewing angle effects \citep{Rossi2002}. If the jet structure is universal (i.e.\  all jets share the same properties) then the LF is entirely determined by viewing angle effects: for a population with isotropic orientations and a monotonic viewing-angle-dependent luminosity $L_\mathrm{iso}(\theta_\mathrm{v})$ (common to all events), the LF can be obtained \cite{Rossi2002,Pescalli2015-bm} by application of the chain rule, namely
\begin{equation}
 \frac{\mathrm{d}R_0}{\mathrm{d}L_\mathrm{iso}}=\frac{\mathrm{d}R_0}{\mathrm{d}\theta_\mathrm{v}}\frac{\mathrm{d}\theta_\mathrm{v}}{\mathrm{d}L_\mathrm{iso}}=R_0\left.\left(\frac{\partial}{\partial \theta_\mathrm{v}}L_\mathrm{iso}(\theta_\mathrm{v})\right)^{-1}\sin\theta_\mathrm{v}\right|_{\theta_\mathrm{v}=f^{-1}(L_\mathrm{iso})},
 \label{eq:LF_universal_SJ}
\end{equation}
where $f^{-1}$ is the inverse of $L_\mathrm{iso}(\theta_\mathrm{v})$, i.e.\ $f^{-1}\left(L_\mathrm{iso}(\theta_\mathrm{v})\right)=\theta_\mathrm{v}$. In the simple case of a power law dependence of the luminosity on the viewing angle, $L_\mathrm{iso}(\theta_\mathrm{v})\propto \theta_\mathrm{v}^{-a}$, and using the small-angle approximation $\sin\theta_\mathrm{v}\sim\theta_\mathrm{v}$, one obtains \cite{Pescalli2015-bm} $\mathrm{d}R_0/\mathrm{d}L_\mathrm{iso}\propto L^{-1-2/a}$. The fact that this asymtpotes to $L^{-1}$ when $a\to\infty$ shows that, due to viewing angle effects, the LF cannot in principle be shallower than $L^{-1}$ due to the contribution of off-axis jets, no matter how suppressed their emission is at large viewing angles\footnote{In practice, the spectrum of a far off-axis jet would be much softer than a typical GRB, the duration much longer, and the light curve smooth \cite{Ascenzi2020}. These features would lead to a different classification than a GRB. As an additional note, the $L^{-1-2/a}$ behaviour breaks down when the small angle approximation $\sin\theta_\mathrm{v}\sim \theta_\mathrm{v}$ becomes invalid. The assessment of the contribution of far off-axis jets to the GRB LF thus requires additional care.}. 

In practice, even in the case in which the progenitor parameter space for which a relativistic jet can be launched is very narrow \respone{(as suggested, for example, by the relatively small spread in the peak luminosities of supernovae associated to long GRBs, e.g.\ \citep{Cano2017,Lu2018})}, some spread in the jet properties within the population is unavoidable. For that reason, even in the wildest unification fantasy the best that can be expected is a \textit{quasi-universal} jet structure with parameters that are spread around a ``typical'' value. The LF in a quasi-universal structured jet scenario can then be seen as a convolution of probability distributions,
\begin{equation}
 P(L_\mathrm{iso})=\int_0^{\pi/2} P(L_\mathrm{iso}\,|\,\theta_\mathrm{v})P(\theta_\mathrm{v})\,\mathrm{d}\theta_\mathrm{v},
\end{equation}
where $P(\theta_\mathrm{v})=\sin\theta_\mathrm{v}$ and $P(L_\mathrm{iso}\,|\,\theta_\mathrm{v})$ is the probability distribution 
of the isotropic-equivalent luminosity at a given viewing angle, which is in turn induced by the probability distributions of the jet structure parameters \cite{Salafia2022_sjpop}.

The black symbols with error bars in the top panel of Fig.~\ref{fig:LF} show the LF of long GRBs as obtained by combining samples of high luminosity (HL) GRBs with measured $z$ and estimated isotropic equivalent luminosities $L_{\rm iso}\geq 10^{50}$ erg/s \cite{Wanderman2010-cb}. The plot also shows the extension of the LF to intermediate luminosities (IL) $10^{48}\leq L_{\rm iso}/(\mathrm{erg\,s^{-1}})<10^{50}$ where, due to selection effects, only lower limits on the intrinsic rate density can be placed \cite{Pescalli2015-bm}, and to the low luminosity range (LL -- $L_{\rm iso}\lesssim 10^{47}$ erg s$^{-1}$) dominated by few events detected in the very local Universe \cite{Pescalli2015-bm}. In the bottom panel of Fig.~\ref{fig:LF}, the green and purple lines and bands show the inverse cumulative LFs of short GRBs obtained by two different studies  \cite{Wanderman2015-ji,Ghirlanda2016-hm} based on the observed properties of the {\it Swift} and {\it Fermi} samples. 

The LF of both long and short GRBs extends over eight orders of magnitudes in luminosity and presents a steep decay $L^{-\alpha}$ with $\alpha\gtrsim 3$ at high luminosities ($L>L_{\rm break}\approx 10^{52}$ erg/s). In the structured jet framework, the most luminous events are likely those observed within the core opening angle. As discussed above, the intermediate and faint end of the LF (for $L<L_{\rm break}$) is at least in part shaped by the jet structure \cite{Pescalli2015-bm}. In long GRBs, the binned LF can be reproduced by a quasi-universal structured jet model \cite{Pescalli2015-bm} with a power law dependence of the luminosity on the viewing angle $L_\mathrm{iso}(\theta_\mathrm{v})=L_\mathrm{c}\min(1,(\theta_\mathrm{v}/\theta_\mathrm{c})^{-a})$ with $a\gtrsim 6$ and $\theta_\mathrm{c}$ and $L_\mathrm{c}$ narrowly distributed around $5^\circ$ and $3\times 10^{52}$ erg/s respectively (cyan solid line in the top panel of Fig.~\ref{fig:LF}). In short GRBs, due to the scarcer data and the absence of an agreement on the general features of the LF, the situation is more unclear. Yet, an attempt at deriving a quasi-universal jet structure model from modelling the jet propagation through, and breakout from, binary neutron star merger ejecta \cite{Salafia2020_jet_propagation} does produce a LF (blue solid line in Fig.\ref{fig:LF}) with a steep decay above a break at $L_\mathrm{break}\sim 3\times 10^{52}$ erg/s and a relatively less steep distribution at lower lumionsities, which gets shallower as it extends to the LL range where the observation of GRB170817A \cite{Abbott2017-mm} places some constraints \cite{Ghirlanda2019}. A shallower (e.g.\ a power law $L_\mathrm{v}\propto \theta_\mathrm{v}^{-2}$) structure would result in a steeper LF that would overproduce the LL long GRBs rate density (blue symbol in Fig.~\ref{fig:LF}, top panel) or, in short GRBs, the binary neutron star merger rate (pink shaded region in Fig.~\ref{fig:LF}, bottom panel). 

\subsection{Jet structure and the $E_\mathrm{peak}-E_\mathrm{iso}$ correlation}

\begin{figure}
    \centering
    \includegraphics{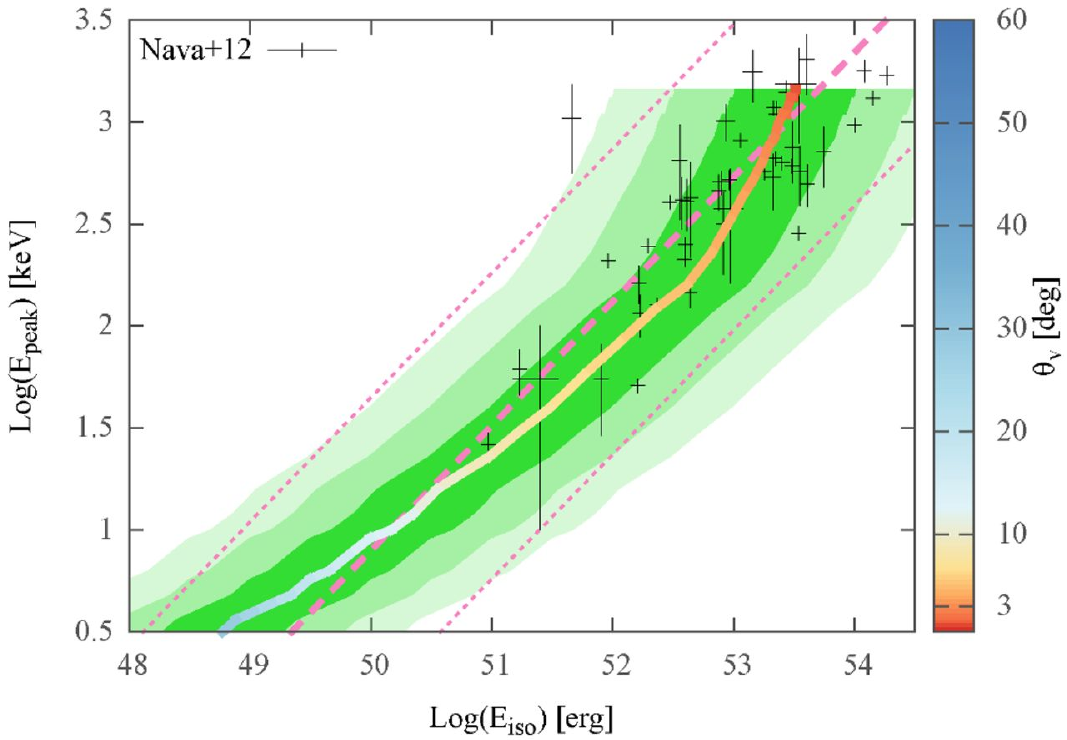}
    \caption{Rest frame peak energy $E_{\rm peak}$ versus isotropic equivalent energy $E_{\rm iso}$ of long GRBs. The data points (cross symbols) are a flux limited sample of bright {\it Swift} bursts. The dashed (dotted) line shows the correlation regression line (and its 3$\sigma$ scatter). The color coded solid line shows the values of $E_{\rm peak}$ and $E_{\rm iso}$ assuming a structured Gaussian jet seen under progressively larger viewing angles -  vertical color-code bar). The green shadows, representing the 1,2,3 $\sigma$ confidence levels around the color--coded line, are obtained considering a dispersion of the core energy of 0.5 dex around a nominal value of $3\times10^{53}$ erg. Figure reproduced from \cite{Salafia2015-dj}.}
    \label{fig:amati}
\end{figure}

A fraction of the GRBs (around one third) that trigger the Burst Alert Telescope (BAT) onboard {\it Swift} end up with a measurement of their redshits \cite{Lien2016}. This allows, in most cases\footnote{The estimate of $E_\mathrm{iso}$ and $L_\mathrm{iso}$ require to measure the SED peak which is often possible thanks to the detection of the burst also by the {\it Fermi} satellite which provides a broad band (10keV-40MeV) energy spectral coverage.}, to estimate rest frame properties such as $E_\mathrm{peak}$, $E_\mathrm{iso}$ and $L_\mathrm{iso}$. 
Fig.~\ref{fig:amati} shows the GRBs from \cite{Nava2012-ny} on the $(E_{\rm iso}, E_{\rm peak})$ plane (black symbols), demonstrating the apparent correlation between these two quantities \cite{Amati2002-qi}. Such a correlation is naturally expected in a quasi-universal structured jet scenario, given the common dependence of $E_\mathrm{iso}$ and $E_\mathrm{peak}$ on the viewing angle  (e.g.\ \cite{Eichler2004}). Assuming Gaussian profiles $\eta_\gamma\mathrm{d}E/\mathrm{d}\Omega=\epsilon_0 \exp(-\theta^2/\theta_\mathrm{c}^2)$ and $\Gamma(\theta)=1 + (\Gamma_\mathrm{c}-1)\exp(-\theta^2/\theta_\mathrm{c}^2)$, an angle-independent comoving peak SED photon energy $E'_\mathrm{peak}=1\,\mathrm{keV}$ \cite{Ghirlanda2012MNRAS.420..483G}, and a quasi-universal structured jet scenario in which the structure parameters in the population are narrowly distributed around typical values $\left\langle\epsilon_0\right\rangle=3\times 10^{53}$, $\left\langle\Gamma_0\right\rangle=800$, $\left\langle\theta_{\rm c}\right\rangle=3^\circ$, the authors of \cite{Salafia2015-dj} could reproduce both the LF of long GRBs and the observed $E_\mathrm{peak}-E_\mathrm{iso}$ correlation, as shown by the model distribution represented in Figure \ref{fig:amati}. The horizontal dispersion in the figure corresponds to just considering a 0.5 dex log-normal dispersion of the core energy density $\epsilon_0$. As shown by the color-coded viewing angle, within this intepretation the known long GRBs are observed within $\theta_\mathrm{v}\lesssim 3\theta_\mathrm{c}\sim 9^\circ$, which is consistent with the constraints derived by \cite{Beniamini2019}. In the bottom left corner, corresponding to $E_{\rm iso}<10^{51}$ erg and $E_{\rm peak}<30$ keV, should reside jets observed at larger viewing angles. If these were detected with next-generation instruments with wide-field, highly sensitive hard X-ray monitors, they could probe the expected bending of the $E_{\rm peak}-E_{\rm iso}$ correlation induced by large viewing angles and in turn help constraining the quasi-universal jet structure scenario.  

\subsection{Jet structure and `late-prompt' emission}\label{sec:lateprompt}

The observed X-ray emission of GRBs extending after the prompt phase up to few hours is often characterized by a steep decay \cite{Tagliaferri2005-yw} of the flux transitioning to a shallow (so called \textit{plateau}) phase \cite{OBrien2006-nl}. On top of this, X--ray flares are often observed \cite{Burrows2005-yk,Nousek2006-qa,Chincarini2010-sh}. Intriguingly, these features of the early X--ray emission can be explained within a structured jet scenario: the steep--plateu shape is what an observer nearly aligned with the jet axis would see, while an off axis observer should see a more uniform powerlaw decay \cite{Oganesyan2020-ka,Ascenzi2020}. In this interpretation, the X--ray light curve up to the end of the plateu phase would have an internal origin, being produced by prompt emission photons reaching the observer from increasingly high-latitude parts of the jet (hence the delayed arrival time, \citep{Kumar2000}). Flares have been explaned, in the context of a structured jet, as late time internal dissipation episodes whose brightness and spectral hardness is reduced by the debeaming effects for an observer with a viewing angle far from the jet axis \cite{Duque2021-rr}.

\section{Afterglow emission from a structured jet}

As explained in \S\ref{sec:stages}, the expansion of the jet material in the ``circum-burst medium'' (the relic stellar wind from the progenitor massive star in long GRBs\footnote{In most long GRBs, the afterglow seems to be best modelled assuming a homogeneous interstellar medium (ISM), which poses a challenge to the massive star progenitor scenario for long GRBs \cite{VanMarle2006}.}, or a tenuous interstellar medium in short GRBs) leads to the formation of a shock, which is widely believed to be the main source of the broad-band GRB afterglow emission \cite{Paczynski1993,Meszaros1997,Sari1998,Panaitescu2000}. Hereon, we focus on the case of a homogeneous ISM composed of hydrogen, with number density $n$, and discuss the impact of jet structure on the expected afterglow emission. 

\subsection{Jet structure and the early aferglow}
\begin{figure}
 \centering
 \includegraphics[width=10cm]{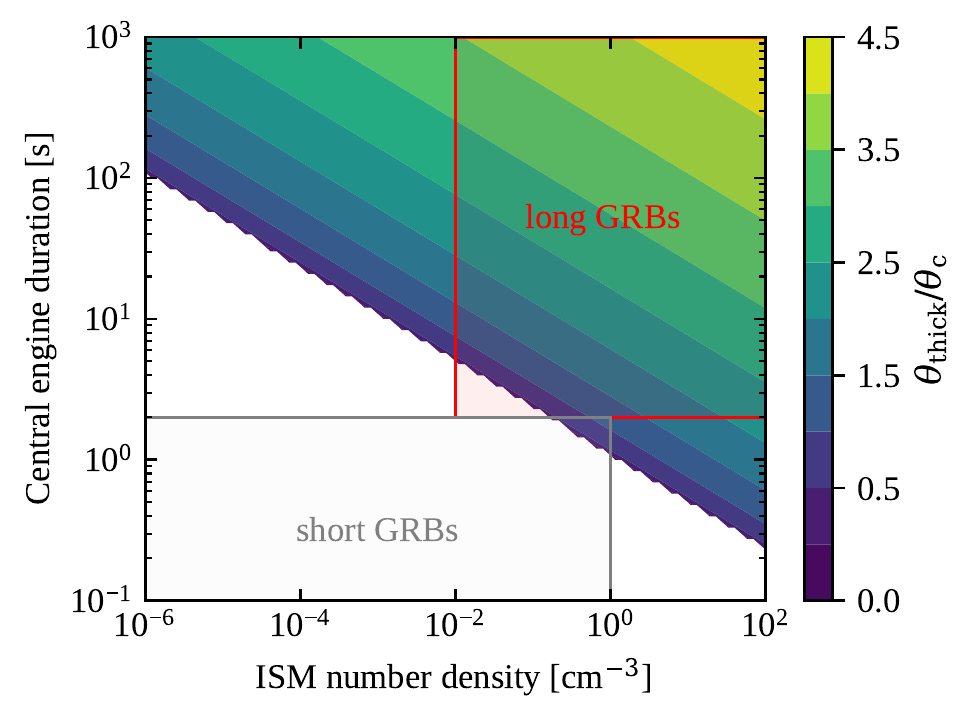}
 \caption{Angle $\theta_\mathrm{thick}$ within which the deceleration takes place in the thick shell regime, in units of the core angle $\theta_\mathrm{c}$, in a structured jet with Gaussian energy $\mathrm{d}E/\mathrm{d}\Omega(\theta)$ and bulk Lorentz factor $\Gamma(\theta)$ profiles $\propto \exp(-(\theta/\theta_\mathrm{c})^2/2)$, assuming $4\pi\mathrm{d}E/\mathrm{d}\Omega(0)=10^{54}$ erg and $\Gamma(0)=1000$, as a function of the ISM number density $n$ and central engine duration $T_\mathrm{CE}$. Boxes show the regions of the plane where most long (red box) and short (grey box) GRBs are expected to lie.}
 \label{fig:RS_thinshell}
\end{figure}

Initially, as the expanding jet material is highly relativistic, there is no causal contact between regions at angular distances $\gtrsim 1/\Gamma$ and hence the shock dynamics only depends on local quantities. Calling $E(\theta)=4\pi\mathrm{d}E/\mathrm{d}\Omega(\theta)$, as soon as the isotropic-equivalent ISM mass swept by the jet, $M(\theta)= n m_\mathrm{p} 4\pi R(\theta)^3/3$ (where $m_\mathrm{p}$ is the proton mass and $R$ is the outer radius of the jet or, more precisely, that of the forward shock) becomes comparable to $E(\theta)/\Gamma^2(\theta)c^2$, a reverse shock starts to propagate backwards (as seen in the rest-frame of the contact discontinuity that separates the shocked ISM and jet materials) through the jet, initiating the deceleration of the latter. Assuming a jet radial width $\Delta_0\sim cT_\mathrm{CE}$ after breakout, the initial deceleration phase proceeds differently depending on the local Sedov length $l_\mathrm{S}(\theta)=(3E(\theta)/4\pi n m_\mathrm{p} c^2)^{1/3}$ and bulk Lorentz factor $\Gamma(\theta)$. Assuming free expansion, the ``sound-crossing'' radius at which radial sound waves (assuming a relativistic sound speed $c_\mathrm{s}=c/\sqrt{3}$) cross the shell is $R_\mathrm{s}(\theta)\sim\sqrt{3}\Gamma(\theta)^2\Delta_0$. The ``deceleration'' radius at which the interaction with the ISM becomes relevant is $R_\mathrm{d}(\theta)\sim l_\mathrm{S}(\theta)\Gamma^{-2/3}(\theta)$. If $R_\mathrm{s}(\theta)>R_\mathrm{d}(\theta)$, the portion of the jet is said to be in the ``thick shell'' regime, where the deceleration starts before radial pressure waves can smooth out radial inhomogeneities and cause any significant radial spreading \cite{Meszaros1993,Sari1995}: in this case, the reverse shock is relativistic (i.e.\ the velocity of the unshocked jet is relativistic as seen from the contact discontinuity that separates the shocked ISM and shocked jet material) and crosses the whole jet shell at a radius $R_\mathrm{cross}(\theta)\sim l_\mathrm{S}(\theta)^{3/4}\Delta_0^{1/4}$. Conversely, if $R_\mathrm{s}(\theta)<R_\mathrm{d}(\theta)$ the jet portion is in the ``thin shell'' regime, where it reaches the deceleration radius after undergoing a significant radial spread, which washes out radial inhomogeneities and leads to an effective jet radial with $\Delta(\theta)\sim R/\Gamma(\theta)^2$. In this case, the reverse shock remains Newtonian and crosses the shell at $R_\mathrm{cross}(\theta)\sim R_\mathrm{d}(\theta)$. Interestingly, as shown in Figure \ref{fig:RS_thinshell}, assuming $E(0)=10^{54}\,\mathrm{erg}$ and $\Gamma(0)=1000$ (the dependence on these values is weak) and adopting a Gaussian profile $\propto \exp(-(\theta/\theta_\mathrm{c})^2/2)$ for both quantities, for most short GRBs (with $T_\mathrm{CE}\lesssim 2\,\mathrm{s}$ and $n\lesssim 1\,\mathrm{cm^{-3}}$) the deceleration is entirely in the thin shell regime (see also \cite{Lamb2019}), while for long GRB jets (typically with $ T_\mathrm{CE}\sim 30\,\mathrm{s}$ and $n\sim 1\,\mathrm{cm^{-3}}$) it proceeds in the thick shell regime within an inner region $\theta<\theta_\mathrm{thick}$ which corresponds typically to the jet core, $\theta_\mathrm{thick}/\theta_\mathrm{c}\sim 1$. More generally, within the above Gaussian structured jet assumption, the existence of a transition angle $\theta_\mathrm{thick}$ corresponds to the condition $\Gamma(0)> (l_\mathrm{S}(0)/\sqrt{3}c T_\mathrm{CE})^{3/8}\sim 430\,E_{54}^{1/8}n_0^{-1/8}T_{\mathrm{CE},1}^{-3/8}$, in which case $\theta_\mathrm{thick}=(2\theta_\mathrm{c}/\sqrt{7})[8\ln\Gamma(0)-3\ln(l_\mathrm{S}(0)/\sqrt{3}cT_\mathrm{CE})]^{1/2}$. 

During this phase, diffusive shock acceleration of electrons \cite{Fermi1949,Bell1978,Blandford1987,Spitkovsky2008} can take place at both the forward and reverse shocks, leading to synchrotron \cite{Kobayashi2000}  and possibly inverse Compton radiation. For viewing angles close to the jet axis, where the emission is dominated by material moving close to the line of sight, the reverse shock emission is expected to peak at the observer time $t_\mathrm{pk,RS}\sim (1+z) R_\mathrm{cross}(\theta_\mathrm{v})/\Gamma(\theta_\mathrm{v})^2 c$ that corresponds to the shock crossing time. In the thin shell regime, this matches the peak time of the forward shock emission. The light curve of the reverse shock emission in the Optical (where the peak of the synchrotron spectrum at $t_\mathrm{pk,RS}$ is expected to lie for ``standard'' parameters, \cite{Kobayashi2000}) and X-rays is expected to display a rapid rise and decay before and after the peak, therefore appearing as a flare. In the radio the expected decay is slower (as the synchrotron peak moves rapidly to lower frequencies after the peak), with possible late-time bumps \cite{Resmi2016}, even though this critically depends on how rapidly the shock-generated magnetic field decays after the reverse shock has disappeared \cite{Salafia2022_29A}. The emission as seen by a far off-axis observer may be instead dominated by material moving at a different angle, or more generally consist of a comparable amount of radiation from a broader portion of the jet, resulting in a delayed and smoother light curve (see \cite{Lamb2019} for examples in short GRBs).

In parts of the jet where the deceleration proceeds in the thick shell regime, the radial structure plays a role in the reverse-forward shock dynamics, and thus in shaping the early afterglow emission. This is particularly relevant in far off-axis parts of the jet that contain blown-out cocoon material (which is expected to feature a broad velocity profile, e.g.\ \cite{Nakar2017,Gottlieb2018,Gottlieb2021}) and/or if the central engine does not turn off abruptly at $T_\mathrm{CE}$, but rather decays slowly, resulting in a jet with a low-velocity tail that contains a non-negligible amount of energy. In that case, the reverse shock can be long-lived, with slower material gradually catching up, with a modified dynamics of both the reverse and forward shocks: this is often called a \textit{refreshed} shock \cite{Rees1998} scenario. This remains currently one of the leading explanations for the X-ray \textit{plateaux} (see \S\ref{sec:lateprompt}). The latter phenomenon is therefore mostly explained as resulting from either a radial or an angular jet structure \cite{Beniamini2020MNRAS.492.2847B}: the degeneracy between these two options when trying to explain ``non-standard'' decays in afterglow light curves has been addressed in \cite{Nakar2018}.

\subsection{Jet structure and the late afterglow}\label{sec:late_afterglow}

As soon as the reverse shock has disappeared, the shocked jet shell transfers most of its energy to the shocked ISM \cite{Kobayashi1999,Kobayashi2000_numerical} and the system turns into a \textit{blastwave}, that is, a forward shock whose memory of the details of the original explosion is lost. In this phase, its dynamics depends solely on the (angle-dependent) Sedov length $l_\mathrm{S}(\theta)\propto (E(\theta)/n)^{1/3}$ and can be described by the evolution of the Lorentz factor of the shocked ISM immediately behind the forward shock, as a function of the forward shock radius (and the angle, given the anisotropy) $\Gamma(R,\theta)$. As long as the blastwave is relativistic, $\Gamma(R,\theta)\gg 1$, its radial structure and evolution are well-described by the self-similar Blandford-McKee solution \cite{Blandford1976}, stably against perturbations \cite{Kobayashi1999}, and hence $\Gamma(R,\theta)=(R/l_\mathrm{S}(\theta))^{-3/2}$ \cite{Blandford1976,Rossi2002,Nava2013}. Lateral energy transfer (which typically proceeds from the jet axis outwards) starts being relevant as soon as transverse causal contact is established, which happens at each angle $\theta$ when $[\Gamma(R,\theta)]^{-1}\gtrsim \theta$. From that radius on, lateral pressure waves will start transferring shock energy laterally, gradually smearing out the angular energy profile $E(\theta)$ \cite{Kumar2003,Granot2003,Rossi2004,Lu2020}. As soon as the entire shocked region reaches transverse causal contact, it expands laterally \cite{Rhoads1999,Cannizzo2004,Zhang2009,Granot2012}, gradually reaching isotropy. At late times, when the whole shock becomes non-relativisic and the expansion quasi-spherical, the shock structure and evolution is well-described \cite{Zhang2009,VanEerten2010,DeColle2012} by the Sedov-Taylor self-similar solution \cite{Taylor1950,Sedov1959,Ostriker1988}. This qualitative description should make clear that most of the memory of the angular structure is wiped out during the lateral spreading phase. 

\respone{The shock dynamics and emission in the structured jet scenario in this phase have been addressed by several studies \citep[e.g.][]{Rossi2002,Kumar2003,Granot2003,Wei2003,Salmonson2003-nf,Rossi2004,Lamb2017-ih,Gill2018,Wu2018,Salafia2019_onaxisview,Beniamini2020-us,Lu2020,Lamb2021-jz,Beniamini2022-yx}}. At any time in the evolution, the emission is typically modelled within a parametrized relativistic leptonic diffusive shock acceleration and synchotron/inverse-Compton emission model \cite{Sari1998,Panaitescu2000}. Shock-accelerated electrons are assumed to be injected in the shock downstream with a power-law distribution of Lorentz factors, $\mathrm{d}n_\mathrm{e}/\mathrm{d}\gamma \propto \gamma^{-p}$ (where $n_\mathrm{e}$ is the comoving electron number density in the immediate downstream, and we focus on the case where $p>2$), above a minimum Lorentz factor $\gamma_\mathrm{m}$. Their number is assumed to be a fraction\footnote{Most often, the fraction is set to $\chi_\mathrm{e}=1$, despite the theoretical expectation being $\chi_\mathrm{e}\sim \mathrm{few}\times 10^{-2}$ \cite{Achterberg2001,Spitkovsky2008}. Yet, in some GRB afterglows with broad-band, high-cadence datasets, $\chi_\mathrm{e}<1$ has been shown to provide a substantially better fit to the data, e.g.\ \cite{Cunningham2020,Salafia2022_29A}.} $\chi_\mathrm{e}$ of the total shocked ISM electrons, and their energy density is assumed to be a fraction $\epsilon_\mathrm{e}$ of the internal energy in the shock downstream (which in a strong shock depends only on the shock velocity/Lorentz factor and on the adiabatic index, being set by shock-jump conditions \cite{Taub1948}): the minimum Lorentz factor $\gamma_\mathrm{m}$ is entirely determined once the shock Lorentz factor $\Gamma$ and the $\chi_\mathrm{e}$, $\epsilon_\mathrm{e}$ and $p$ parameters are given. A random mangetic field generated by small-scale turbulence behind the shock \cite{Medvedev1999} is assumed to hold a fraction $\epsilon_\mathrm{B}$ of the energy density. The electron population in the shock downstream evolves as fresh electrons are injected, and as older electrons cool down due to synchotron, inverse-Compton and adiabatic losses: as a result, the electron distribution in phase space takes approximately the form \cite{Panaitescu2000}
\begin{equation}
 \frac{\mathrm{d}n_\mathrm{e}}{\mathrm{d}\gamma}\propto\left\lbrace
 \begin{array}{lc}
\gamma^{-q} & \gamma_\mathrm{p} \leq \gamma < \gamma_\mathrm{0} \\
\gamma^{-p-1} & \gamma \geq \gamma_\mathrm{0}\\
 \end{array}
\right.,
\end{equation}
where $\gamma_\mathrm{p}=\min(\gamma_\mathrm{m},\gamma_\mathrm{c})$, $\gamma_\mathrm{0}=\max(\gamma_\mathrm{m},\gamma_\mathrm{c})$, 
\begin{equation}
q=\left\lbrace
 \begin{array}{lc}
p & \gamma_\mathrm{m} \leq \gamma_\mathrm{c} \\
2 & \gamma_\mathrm{m}  > \gamma_\mathrm{c} \\
 \end{array}
\right., 
\end{equation}
and $\gamma_\mathrm{c}$ is the Lorentz factor above which electrons loose most of their energy through synchrotron, inverse-Compton and adiabatic losses in a dynamical time $t'_\mathrm{dyn}\sim R/\Gamma c$. This is typically given by \cite{Sari1998,Panaitescu2000}
\begin{equation}
 \gamma_\mathrm{c}=\frac{6\pi m_\mathrm{e}c^2 \Gamma}{\sigma_\mathrm{T}B^2 R (1+Y)},
\end{equation}
where $Y=u_\mathrm{rad}/u_\mathrm{B}$ is the ratio of radiation energy density\footnote{This includes the synchrotron radiation produced by the electrons -- which emit and cool both by synchrotron and by inverse-Compton scattering of their own synchrotron photons, i.e.\ synchrotron-self-Compton \cite{Dermer2000,Sari2001} -- and possibly an external radiation field, which can be upscattered by relativistic electrons in the shock downstream giving rise to an additional emission component \cite{Ghisellini1991,Fan2006,Mei2022}.} to magnetic energy density as measured in the shock downstream comoving frame.
More complicated electron energy distributions arise when Klein-Nishina effects are important and most electrons cool rapidly, e.g.\ \cite{Nakar2009,Daigne2011}. 

The above simple model of electron acceleration and cooling thus depends on the ``microphysical'' parameters $p$, $\chi_\mathrm{e}$, $\epsilon_\mathrm{e}$ and $\epsilon_\mathrm{B}$, and on the shock Lorentz factor $\Gamma$. Once the shock dynamics $\Gamma(R,\theta)$ is determined, the luminosity emitted towards an observer at a viewing angle $\theta_\mathrm{v}$ from electrons in the entire shock can thus be computed for a fixed set of microphysical parameters by integrating the radiative transfer equation after computing the synchrotron (and possibly inverse-Compton) emissivity (and absorption coefficient) over the shock downstream, the shock radial structure being given by the appropriate self-similar solution. In order for the resulting luminosity to be appropriately compared to the observed flux, the integration must be done over the appropriate equal-arrival-time volume or, in other words, at the appropriate `retarded' time to account for the different photon paths that lead to the same arrival time to the observer \cite{Rees1966,Sari1998b,Granot1999,Granot2002,Ghirlanda2019}. If the shock is approximated as infinitely thin, this reduces to equal-arrival-time surfaces \cite{Panaitescu1998,Panaitescu2000,Rossi2004,Salafia2019_onaxisview}.

\begin{figure}
    \centering
    \includegraphics[scale=1.4]{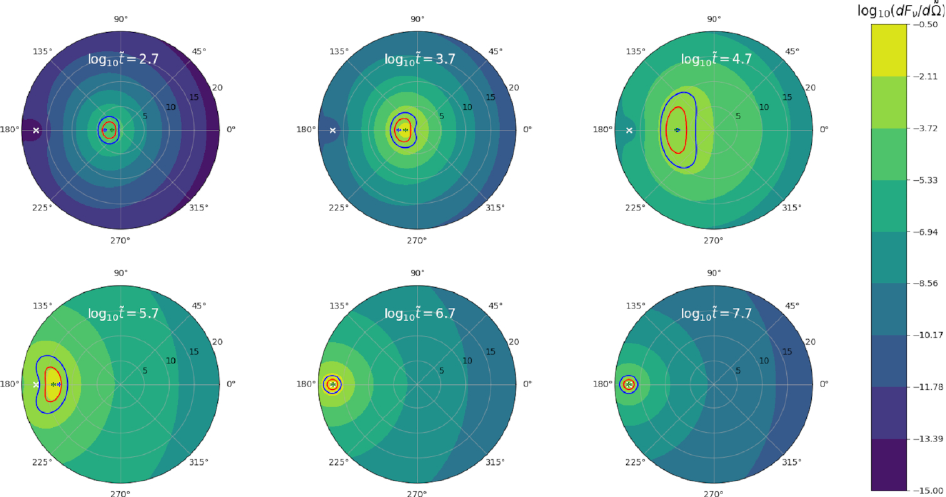} 
    \caption{Angular map showing the intensity distribution per unit solid angle (color coded) of emission from the afterglow forward shock, centered on the line of sight. The position of the jet axis is marked by the white cross symbol. The green cross and red (blue) contours shows the peak of the intensity and the region containing 50 (80\%) of the total flux. A power-law structured jet with core values $\Gamma_\mathrm{c}=1000$ and $\theta_\mathrm{c}=0.03$ rad is considered. The power-law slopes are $a = 4$ and $b = 2$ for the energy and $\Gamma$ structure, the observer viewing angle is 10 times the core opening angle and the external medium density is constant. The maps correspond to different observing times (as measured in the rest frame of the central engine), from 500 s (top left) to 580 days (bottom right). Reproduced from \cite{Beniamini2022-yx}.}
    \label{fig:images}
\end{figure}

Even when limiting the discussion to an isotropic explosion and to synchrotron emission only, the above model leads to rather complex light curves \cite{Granot2002}. At any fixed time, the SED is composed of various smoothly-connected power law segments, corresponding to different spectral regimes: the main critical frequencies are the synchrotron frequencies $\nu_\mathrm{syn}(\gamma)=\delta\gamma^2 eB/2\pi m_\mathrm{e}c$ (where $\delta$ is the Doppler factor related to bulk motion) corresponding to the electron distribution break Lorentz factors $\gamma_\mathrm{m}$ and $\gamma_\mathrm{c}$ (typically referred to as $\nu_\mathrm{m}=\nu_\mathrm{syn}(\gamma_\mathrm{m})$ and $\nu_\mathrm{c}=\nu_\mathrm{syn}(\gamma_\mathrm{c})$), and frequency $\nu_\mathrm{a}$ below which synchrotron self-absorption becomes important (i.e.\ the synchrotron self-absorption optical depth $\tau_\mathrm{ssa}(\nu_\mathrm{a})=1$). The comoving specific synchrotron emissivity at a comoving frequency $\nu'$ can be approximated by a series of power law segments, namely
\begin{equation}
  j'_{\nu'} = j'_{\nu',\mathrm{max}}\left\lbrace
 \begin{array}{lc}
 (\nu'/\nu'_\mathrm{p})^{1/3} & \nu' \leq \nu'_\mathrm{p} \\
 (\nu'/\nu'_\mathrm{p})^{-(q-1)/2} & \nu'_\mathrm{p} < \nu' \leq \nu'_\mathrm{0} \\
 (\nu'_\mathrm{0}/\nu'_\mathrm{p})^{-(q-1)/2}(\nu'/\nu'_\mathrm{0})^{-p/2} & \nu' > \nu'_\mathrm{0} \\
 \end{array}
 \right., 
\end{equation}
where $\nu_0=\nu_\mathrm{syn}(\gamma_0)$,  $\nu_\mathrm{p}=\nu_\mathrm{syn}(\gamma_\mathrm{p})$, and $q$, $\gamma_0$ and $\gamma_\mathrm{p}$ have the same meaning as before. The maximum specific emissivity $j'_{\nu',\mathrm{max}}$ depends on the microphysical parameters, ISM density and shock Lorentz factor  \cite{Sari1998,Salafia2019_onaxisview}. When considering optically thin portions of the shock, the intensity received by the observer is just $I_\nu = \delta^3 j'_{\nu'}(\nu/\delta) \Delta R'$, where $\Delta R'$ is the shock effective thickness, typically of order $\Delta R'\sim R/\Gamma^2$. Hence, if the emission is dominated by a small, optically thin portion of the shock, the observed flux density has the same shape as the emissivity. Synchrotron self-absorption \cite{Rybicki1979}  suppresses the emission when $\tau_\mathrm{ssa}>1$, introducing low-frequency power law segments with $I_\nu \propto \nu^\alpha$ with $\alpha = 2$ if $\nu<\nu_\mathrm{p}$ and $\alpha=5/2$ otherwise. If $\gamma_\mathrm{c}<\gamma_\mathrm{m}$, an additional power law segment with $\alpha=11/8$ emerges below $\nu_\mathrm{a}$ due to the inhomogeneous cooling stage of electrons behind the shock \cite{Granot2000}. Useful figures summarizing all the possible synchrotron spectral regimes in GRB afterglows can be found in \cite{Granot2002}. Reference \cite{Nakar2009} treats additional cases where the electron distirbution (and hence the synchrotron spectrum) is modified by a non-monotonic dependence of the cooling rate on the electron Lorentz factor $\gamma$ due to Klein-Nishina effects. 

\begin{figure}
    \centering
    \includegraphics[width=0.6\textwidth]{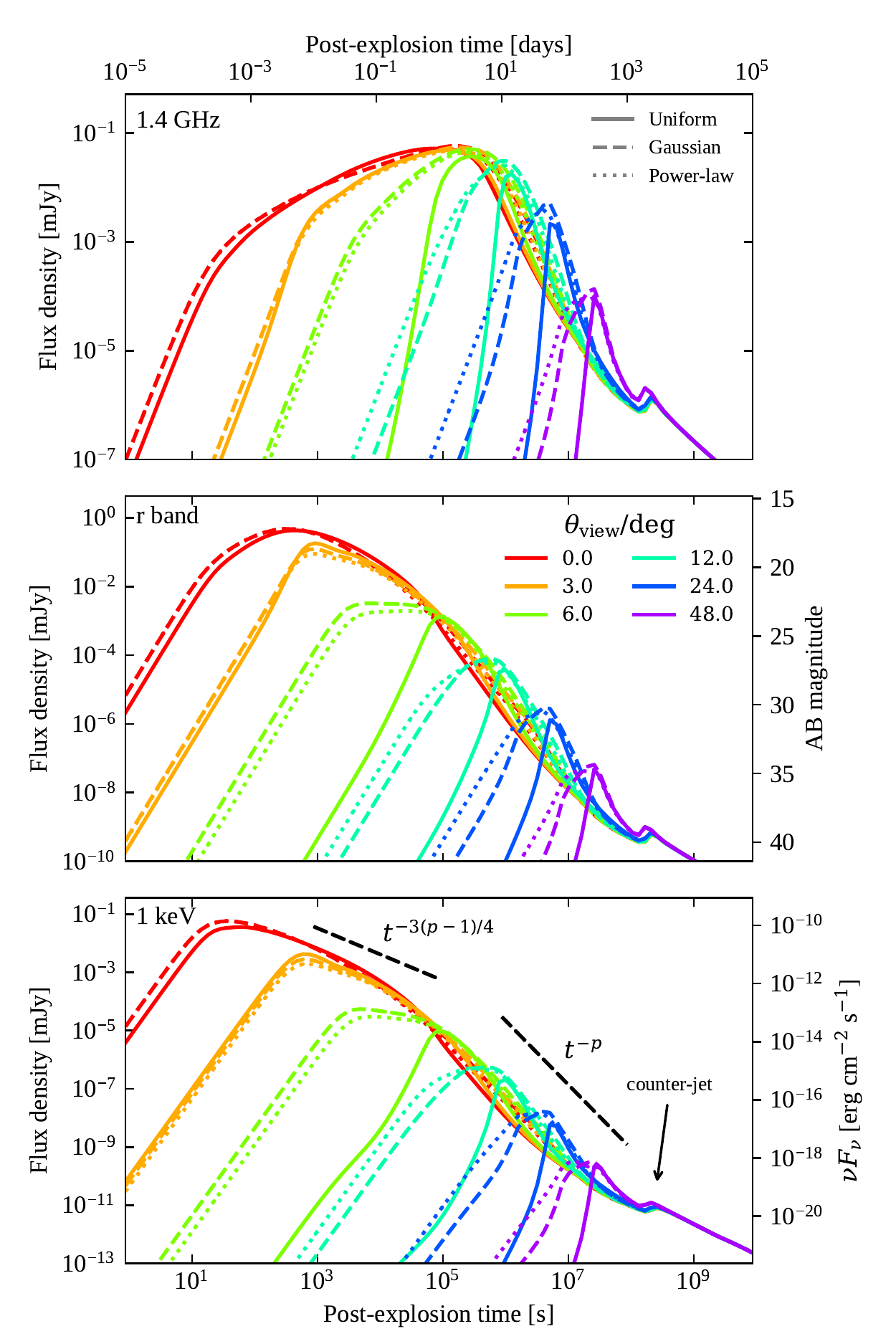}
    \caption{Example synthetic afterglow light curves for jets at $z=1$ with differing structures, seen under various viewing angles (color-coded as reported in the legend), computed within the ``standard afterglow'' model described in \S\ref{sec:late_afterglow} assuming $\epsilon_\mathrm{e}=0.1$, $\epsilon_\mathrm{B}=10^{-3}$, $p=2.2$, and a homogeneous external medium with number density $n=0.1\,\mathrm{cm^{-3}}$. Each structure can be defined as $\mathrm{d}E/\mathrm{d}\Omega = (E_\mathrm{c}/4\pi )f(\theta/\theta_\mathrm{c})$ and $\Gamma = 1 + (\Gamma_\mathrm{c}-1)f(\theta/\theta_\mathrm{c})$: (i) the Uniform structure (solid lines) has $f(x)=\Theta(1-x)$, where $\Theta(x)$ is the Heaviside step function; (ii) Gaussian jet (dashed lines), with $f(x)=\exp(-x^2/2)$; (iii) Power-law jet, with $f(x)=(1+x^3)^{-1}$. For all jets, $E_\mathrm{c}=10^{53}\,\mathrm{erg}$, $\Gamma_\mathrm{c}=300$ and $\theta_\mathrm{c}=3^\circ$. Each panel shows light curves computed at a different observing frequency: 1.4 GHz (top panel), r-band (i.e.\ $4.6\times 10^{14}\,\mathrm{Hz}$, central panel) and 1 keV (i.e.\ $2.4\times 10^{17}\,\mathrm{Hz}$, bottom panel). In the bottom panel, we show for comparison the slopes expected for the X-ray (mostly also valid for the Optical) post-peak (for viewing angles close to the jet axis) and post-jet-break light curves. The late-time peak produced by the emission of the counter-jet is also annotated.}
    \label{fig:offaxis-sj-examples}  
\end{figure}

For a given jet structure, observed off the jet core ($\theta_{obs}>\theta_{c}$), the emission is dominated, at different times, by different portions of the emitting surface \cite{Beniamini2020-us,Beniamini2022-yx}. Fig.~\ref{fig:images} shows that as time increases, the flux seen by the observer is dominated by emission produced progressively close to the jet axis as a consequence of the competition between the decrease of the shock velocity (implying the increase of the beaming angle and the decrease of the emissivity) and the increasing shock initial energy towards the jet axis with respect to the observer line of sight (center of the coordinate system in Fig.~\ref{fig:images}). This effect has important consequences on the observed afterglow light curves for different jet structures and viewing angles. Fig.~\ref{fig:offaxis-sj-examples} shows some examples of mono-chromatic afterglow light curves at three characteristic frequencies corresponding to the radio, optical and X-ray bands (from top to bottom). Three different jet structures are considered here: uniform (solid lines), Gaussian (dashed line) power-law with both the energy and the Lorentz factor decreasing as $\theta^{-3}$ (dotted line). In all three cases, a jet core opening angle of 3 degrees is considered. As long as the observer line of sight is within the beaming cone of the jet core, $\theta_\mathrm{v}-\theta_\mathrm{c}<1/\Gamma(R,0)$, the light curves corresponding to different structures are almost indistinguishable. This applies to the entire light curve as long as $\theta_\mathrm{v}<\theta_\mathrm{c}$. If the viewing angle is larger, differences among the three structures are apparent mainly in the rising phases of the light curves, with the two non-uniform structures presenting similar shallow rising phases preceding the peak, after which all structures join into the same, core-dominated decay. \respone{The enormous difference in the light curves at intermediate viewing angles $\theta_\mathrm{c}<\theta_\mathrm{v}\lesssim \mathrm{few}\times \theta_\mathrm{c}$ stems from the fact that, in the Gaussian and power-law structure cases, the emission is initially dominated by material moving along the line of sight, which is absent in the uniform jet case.}

\respone{The fact that the emission is dominated by material progressively closer to the jet axis impacts also the apparent displacement of the source centroid as seen in Very Long Baseline Interferometry (VLBI) imaging \cite{Gill2018,Granot2018a,Granot2018b,Ghirlanda2019,Hotokezaka2019,Fernandez2022}. The surface brightness of the shock, as seen by a distant observer, corresponds to the intensity $I_\nu$ described above. Its distribution $I_\nu(\theta_x,\theta_y)$ on the plane of the sky (where $\theta_x$ and $\theta_y$ represent two suitably chosen angular coordinates on the relevant sky patch) is most commonly referred to as the `image' of the source\footnote{See Figure \ref{fig:jet_cocoon_imgs} in the next section for some example surface brightness distributions.}, which can be measured through VLBI imaging within the limited resolution that can be reached with current facilities. The image centroid (i.e.\ the mean of the distribution) lies on the projection of the jet axis on the $(\theta_x,\theta_y)$ plane, because the jet axisymmetry induces a reflection symmetry in the image\footnote{Such symmetry can also be exploited to speed up the computation of $I_\nu$ for the calculation of light curves \citep{Salafia2019_onaxisview}.}. The displacement of the centroid before the light curve peak (after which the emission becomes core-dominated) is directly related to the shape of the jet structure.}

\respone{The fact that the evolution of the pre-peak light curves and in principle also that of the image centroid position contain some information on the jet structure suggests that it is possible to reconstruct (at least partially) the latter information from the observations. This is typically done by fitting an analytical structured jet afterglow model to the observed light curves (and centroid displacement), in order to recover the parameters that describe the structure \citep[e.g.][]{Gill2018,Resmi2018-kj,Ghirlanda2019,Hotokezaka2019,Lin2019,Lamb2019_GW170817}. As demonstrated by \citep{Takahashi2020}, the light-curve-based reconstruction can be done more explicitly by integrating a differential equation derived from standard afterglow theory. Unfortunately \cite{Takahashi2020,Takahashi2021}, the accuracy and cadence required for a detailed reconstruction are highly demanding, and global degeneracies remain unbroken unless the evolution of the emission in multiple spectral regimes is observed \citep[see also][]{Beniamini2022-yx}.}

\section{GW170817 and GRB170817A: an observational test-bed for off-axis structured jet theory}
\label{sec:GW170817}
 
\begin{figure}
 \centering
 \includegraphics[width=0.6\textwidth]{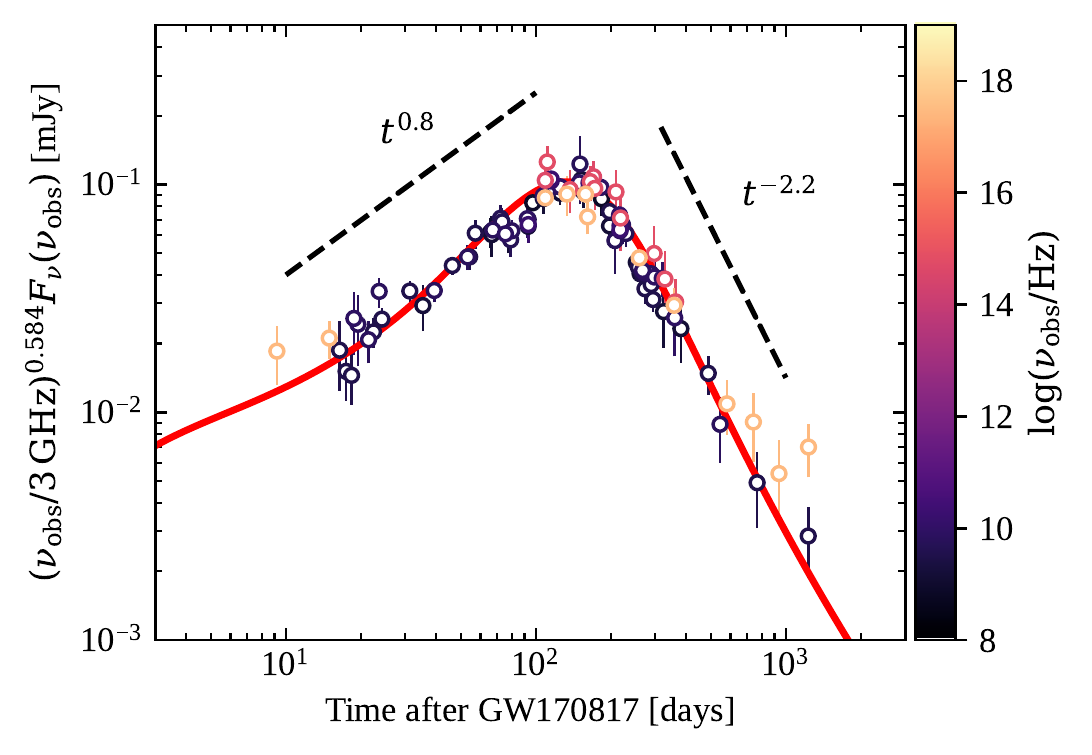}
 \caption{Observations and model of the afterglow of GRB170817A. Colored circles with error bars show flux densities measured in radio, X-ray and optical observations (data from \citep{Makhathini2021}) at the position of GW170817, rescaled to a common frequency of 3 GHz assuming a power-law spectrum $F_\nu\propto \nu^{-0.584}$ \citep{Margutti2018-qf}. The datapoints are color-coded according to the colorbar on the right in order to show the original observing frequency. The red solid line is the prediction of an off-axis structured jet model with a power-law profile of both the energy and the Lorentz factor, with the same form and similar parameters as that in \citep{Ghirlanda2019}, namely $E(\theta)=E(0)/(1+(\theta/\theta_\mathrm{c})^{s_{E}})$ and $\Gamma(\theta)=1 + (\Gamma(0)-1)/(1+(\theta/\theta_\mathrm{c})^{s_{\Gamma}})$ with $\theta_\mathrm{c}=3.2^\circ$, $E(0)=4\times 10^{52}\,\mathrm{erg}$, $s_E=4.5$, $\Gamma(0)=1000$, $s_\Gamma=3.3$, $n=10^{-3}\,\mathrm{cm^{-3}}$, $\epsilon_\mathrm{e}=0.1$, $\epsilon_B=10^{-4}$, $p=2.168$ and $\theta_\mathrm{v}=19^\circ$. }
 \label{fig:17a_afterglow}
\end{figure}

\begin{figure}
 \centering
\includegraphics[height=0.5\textheight]{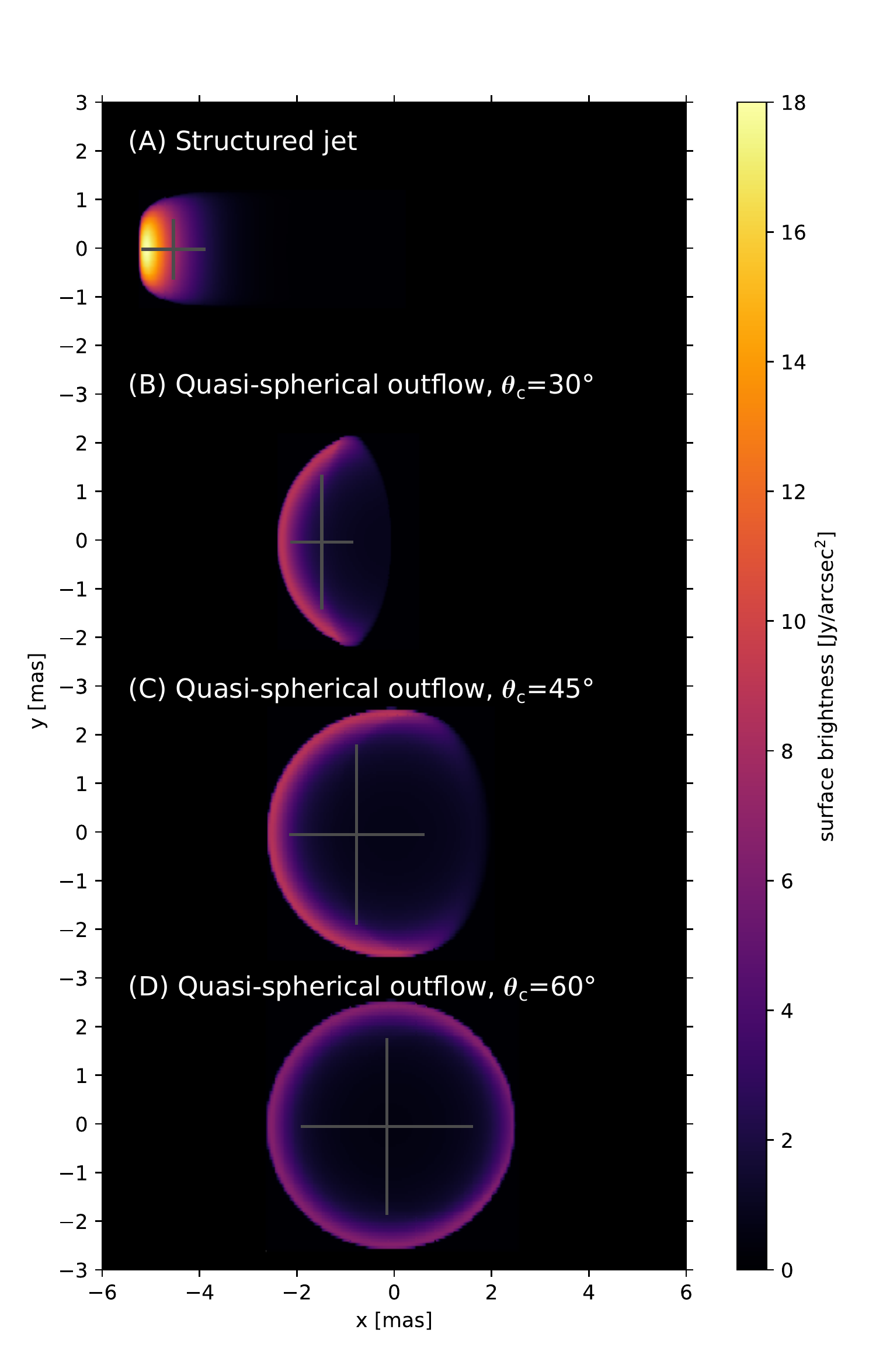}
\caption{Model images of an off-axis structured jet (A) and of three quasi-spherical explosions (B, C and D) with similar parameters, but differing opening angles $\theta_\mathrm{c}$. The angle between the line of sight and the shock symmetry axis in all cases is $\theta_\mathrm{v}=15^\circ$. The images show the surface brightness at 5 GHz, 207 days after the merger, as seen from a distance of 41 Mpc. In each panel, the merger is located at coordinates $(0,0)$, while the grey cross shows the image centroid and the full width at half maximum of the image in two perpendicular directions. All models are compatible with the observations of the GW170817 non-thermal multi-wavelength counterpart (shown in Fig.~\ref{fig:17a_afterglow}) before peak. Adapted from \cite{Ghirlanda2019}. }
\label{fig:jet_cocoon_imgs}
\end{figure}

\begin{figure}
 \centering
\includegraphics[height=0.3\textheight]{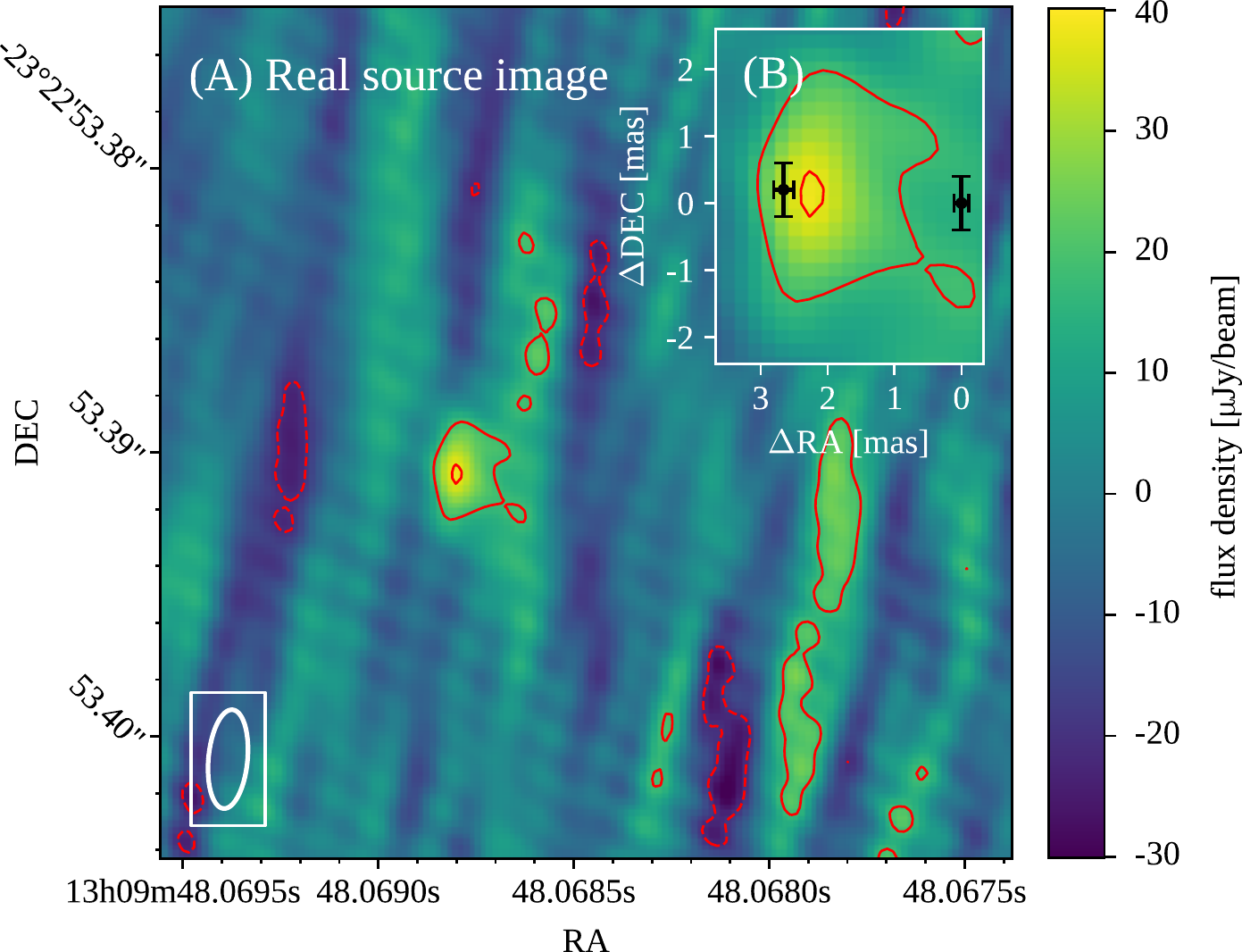}
\caption{Global VLBI image of the GRB170817A afterglow 207 days after the merger. The main panel (A) shows the cleaned surface brightness map (color-coded according to the colorbar on the right) of a small region around the position of GW170817, observed 207 days after the merger. Red contours are lines of constant surface brightness corresponding to -20 (dashed), 20 and 40 (solid) $\mu$Jy/beam. The root-mean-square of the image noise is 8 $\mu$Jy/beam. The ellipse in the lower left corner of the plot encompassess the full width at half maximum of the synthesized beam (i.e.\ the resolution element). The inset, panel (B), shows a zoom of the region close to the peak of the surface brightness distribution, with black error bars marking the best fit and one-sigma errors on the centroid position of the source at 75 and 230 days, as measured by \citep{Mooley2018-xr}, with the axes showing the displacement with respect to the position at 75 d.  Reproduced from \cite{Ghirlanda2019}. }
\label{fig:VLBI_img}
\end{figure}

As noted in the introduction, the discovery of GW170817 and its electromagnetic counterparts \citep{Abbott2017ApJ,Abbott2017PRL} marks a discontinuity in the evolution of the interest of the astrophysical community in structured jets. The reason is that a structured jet observed off-axis provides the most satisfactory and self-consistent explanation for the behavior of the associated short gamma-ray burst GRB170817A \citep{Goldstein2017,Savchenko2017} and of the non-thermal emission observed at the source position starting on the second week after the gravitational wave event in radio \citep{Hallinan2017,Alexander2017,Dobie2018,Resmi2018-kj,Corsi2018,Mooley2018cocoon,Mooley2018-xr,Ghirlanda2019,Balasubramanian2021}, X-rays \citep{Margutti2017,Haggard2017,Troja2017,Nynka2018,Margutti2018-qf,DAvanzo2018-ex,Piro2019,Hajela2019} and optical around peak (with the Hubble Space Telescope \citep{Lyman2018,Fong2019} and with the Large Binocular Telescope \citep{Ghirlanda2019}). Striking evidence in favor of such scenario came from the latter component: its initial light curve evolution, with an unprecedented shallow increase in flux as $\sim t^{0.8}$ over three months (see Fig.~\ref{fig:17a_afterglow}), sparked a debate within the community about its interpretation. The two main competing scenarios attributed the emission to a mildly relativistic shock propagating into the interstellar medium. In one scenario, the shock was produced by an off-axis structured jet \citep[e.g.][]{Lazzati2018-vl,DAvanzo2018-ex} that successfully broke out of the merger ejecta. In the other, it was due to a quasi-spherical outflow with a velocity profile, with most energy in the slower ejecta (models of the radio surace brightness in the two scenarios are shown in Figure \ref{fig:jet_cocoon_imgs}).
The latter outflow could have been either the result of the jet being present, but `choked' \citep{Kasliwal2017,Mooley2018cocoon} (i.e.\ the central engine turned off before the head was able to break out), or could have arised from rapid conversion of magnetic energy into kinetic energy soon after the merger \citep[e.g.][]{Salafia2018,Nathanail2019}. Unfortunately, Nakar \& Piran \citep{Nakar2018} showed that it was impossible to tell apart the two scenarios solely from the light curve evolution before the peak, because a shallow power-law increase in the radio and X-ray flux density could be produced by ejecta with an appropriately chosen angular profile, or velocity profile (or an infinite family of combinations of the two), and in neither case the required parameters were unrealistic. The solution to the riddle was eventually provided by high-resolution VLBI observations \citep{Mooley2018-xr,Ghirlanda2019} at 75, 207 and 230 days after the merger, which revealed an apparently superluminal motion of the radio source centroid and a very small projected size of the image \citep[][see Figure \ref{fig:VLBI_img}]{Ghirlanda2019}. Only the off-axis structured jet scenario has been demonstrated to provide a complete, self-consistent explanation of the light curves and centroid motion to date.

The off-axis viewing angle \citep[most likely in the $15^\circ-25^\circ$ range, see][]{Mooley2018-xr,Nakar2021}, which is in good agreement with the binary orbital plane inclination derived from the gravitational wave analysis \citep{Abbott2017PRL}, has been also identified as the culprit of the extremely low luminosity of the GRB170817A prompt emission \citep[approximately $L_\mathrm{iso}\sim 10^{47}$ erg/s, see][]{Abbott2017ApJ,Goldstein2017,Savchenko2017} when compared to the other known short GRBs with a redshift measurement, since the latter are instead observed within $\sim 2\theta_\mathrm{c}$ \citep{Beniamini2019,Matsumoto2019b}. However, the simple interpretation of GRB170817A as being a regular short GRBs with a suppressed flux due to relativistic beaming \citep[e.g.][]{Pian2017,Abbott2017ApJ} is not viable. Compactness limits \citep{Matsumoto2019a,Matsumoto2019b} indicate that the GRB170817A emission was produced by material moving at a Lorentz factor $\Gamma\gtrsim\mathrm{few}$, but seen under a viewing angle $\theta_\mathrm{e}\lesssim 5^\circ$. Given the viewing angle $\theta_\mathrm{v}\gtrsim 15^\circ$ and the opening angle $\theta_\mathrm{c}\lesssim 5^\circ$ \citep[e.g.][]{Nakar2021}, this is not compatible with emission originating at the border of the jet core, for which $\theta_\mathrm{e}=(\theta_\mathrm{v}-\theta_\mathrm{c})>5^\circ$. The mechanism that produced the observed emission could still have been a similar one as that behind the known short GRB population, but operating well outside the jet core \citep[e.g.][]{Ioka2019}, or a different mechanism, such as the cocoon shock breakout \citep[e.g.][see \S\ref{sec:breakout}]{Kasliwal2017,Gottlieb2018,Nakar2018}.

\section{Concluding remarks}

In this review we attempted at summarizing the historical development of the current ideas on the processes that shape the structure of relativistic jets in gamma-ray bursts and the main observational consequences of that structure. We put some emphasis on the qualitative physical description of the structure formation and on the imprint of the structure on the main observables that characterize the prompt and afterglow emission of gamma-ray bursts, with the aim of providing a first guidance (with a minimal use of technicalities and a pedagogical approach) to those who encounter these topics for the first time. Inevitably, the review only covers a fraction of the relevant literature, only scratches the surface of most arguments, and represents only a partial account of the huge scientific effort performed by the community in the last few decades on this topic. For example, we entirely neglected the important topic of the polarization of the emission from a structured jet: fortunately, this is brilliantly covered by another review article \cite{Gill2022} in this same Special Issue.

Most of the topics presented here are fields of active research and are evolving at a fast pace, especially in the latest years after GW170817. The upcoming science runs of the ground-based gravitational wave detector network, currently comprising the Laser Interferometer Gravitational wave Observatory (LIGO, \cite{Aasi2015}), Advanced Virgo \cite{Acernese2015} and KAGRA \cite{Somiya2012}, will likely soon yield at least one new binary neutron star and/or black hole - neutron star merger event (e.g.\ \cite{Abbott2020_prospects,Abbott2022_grbsearch,Petrov2022,Colombo2022,Howell2019,Mochkovitch2021,Duque2019,Saleem2020,Zhu2021,Yu2021}), hopefully with an associated jet: this will provide new unique insights on the structure of short GRB jets, on the properties of off-axis jet emission in general, and on the incidence of jets, which can be used to constrain the progenitor population (e.g.\ \cite{Salafia2022}). \respone{With new `golden' events such as GW170817, direct information on the jet structure can be extracted, e.g.\ through the methodologies discussed at the end of \S\ref{sec:late_afterglow}, provided that a detailed, high-cadence, multi-wavelength dataset will be collected. The availability of VLBI observations will greatly enhance the chances to break the inherent degeneracies in the afterglow modelling and hence in the jet structure. Moreover, new events observed at differing viewing angles will be the perfect route to test the quasi-universal structured jet hypothesis.} We look forward to learning much more about the structure of gamma-ray burst jets from these and other observations, and from theoretical advances, in the close future.

\acknowledgments{We thank G.\ Ghisellini for years of fruitful discussions. We thank Yuri Sato for useful comments. We thank the anonymous referees for their comments, which helped in improving the quality and completeness of this review. PRIN-INAF “Towards the SKA and CTA era: discovery, localisation, and physics of transient sources” (1.05.01.88.06) and  PRIN-MUR 2017 (grant
20179ZF5KS) are acknowledged for financial support. 
}

\conflictsofinterest{The authors declare no conflict of interest.}

\abbreviations{The following abbreviations are used in this manuscript:\\

\noindent
\begin{tabular}{@{}ll}
GRB Gamma Ray Burst\\
NS Neutron Star\\
BH Black Hole\\
AGN Active Galactic Nucleus\\
CGRO Compton Gamma Ray Observatory\\
BATSE Burst And Transient Source Experiment\\
GBM Gamma-ray Burst Monitor\\
BAT Burst Alert Telescope\\
XRT X-Ray Telescope\\
SSC Synchrotron Self Compton\\
LAT Large Area Telescope
\end{tabular}}

\appendixtitles{no} 
\end{paracol}
\reftitle{References}

\externalbibliography{yes}
\bibliography{references}

\end{document}